\documentclass[journal]{IEEEtran}

\usepackage{amsmath,amssymb,amsfonts,graphicx,color}
\usepackage{mathtools}
\usepackage{commath,bm}

\usepackage{lipsum}
\usepackage{cuted}

\usepackage{mathrsfs} 
\usepackage{cite}
\allowdisplaybreaks

\usepackage{algorithm}
\usepackage{algpseudocode}
\usepackage{array}

\usepackage[caption=false,font=footnotesize]{subfig}

\usepackage{url} 

\begin{document}
\title{Cooperative Beamforming with Predictive Relay Selection for Urban mmWave  Communications}


\author{Anastasios~Dimas,~\IEEEmembership{Student Member,~IEEE,}
        Dionysios~S.~Kalogerias,~\IEEEmembership{Member,~IEEE,}
        and~Athina~P.~Petropulu,~\IEEEmembership{Fellow,~IEEE}
 \thanks{A. Dimas and A. P. Petropulu are with the Department of Electrical and Computer Engineering, Rutgers, The State University of New Jersey, Piscataway, NJ 08854 USA (e-mail: tasos.dimas@rutgers.edu; athinap@rutgers.edu).}
\thanks{D. S. Kalogerias is with the Department of Electrical and Systems Engineering, University of Pennsylvania, Philadelphia, PA 19104, USA (e-mail: dionysis@seas.upenn.edu)}
\thanks{This work was supported by the National Science Foundation under Grant CCF-1526908.}}


%
%

\markboth{Journal of \LaTeX\ Class Files}%
{Shell \MakeLowercase{\textit{et al.}}: Bare Demo of IEEEtran.cls for IEEE Journals}


\maketitle

\begin{abstract}
While millimeter wave (mmWave) communications promise high data rates, their sensitivity to blockage and severe signal attenuation presents challenges in their deployment in urban settings. To overcome these effects, we consider a distributed cooperative beamforming system, which relies on static relays deployed in clusters with similar channel characteristics, and where, at every time instance, only one relay from each cluster is selected to participate in beamforming to the destination. To meet the quality-of-service guarantees of the network, a key prerequisite for beamforming is relay selection. However, as the channels change with time, relay selection becomes a resource demanding task. Indeed, estimation of channel state information for all candidate relays, essential for relay selection, is a process that takes up bandwidth, wastes power and introduces latency and interference in the network. We instead propose a unique, predictive scheme for resource efficient relay selection, which exploits the special propagation patterns of the mmWave medium, and can be executed distributively across clusters, and in parallel to optimal beamforming-based communication. The proposed predictive scheme efficiently exploits spatiotemporal channel correlations with current and past networkwide Received Signal Strength (RSS), the latter being invariant to relay cluster size, measured sequentially during the operation of the system. 
Our numerical results confirm that our proposed relay selection strategy outperforms any randomized selection policy that does not exploit channel correlations, whereas, at the same time, it performs very close to an ideal scheme that uses complete, cluster size dependent RSS, and offers significant savings in terms of channel estimation overhead, providing substantially better network utilization, especially in dense topologies, typical in mmWave networks.
\end{abstract}

\begin{IEEEkeywords}
mmWave, cooperative relay beamforming, relay selection, $2$-hop relaying, mmWave channel modeling, Vehicle-to-Infrastructure.
\end{IEEEkeywords}

%
\IEEEpeerreviewmaketitle


\section{Introduction}
\label{sect:introduction}
\IEEEPARstart{T}{he} continuously growing number of connected devices has led to congestion in the licensed spectrum. To alleviate the problem,  next generation of commercial wireless networks will exploit the previously untapped millimeter wave (mmWave) spectrum band \cite{federal2016spectrum}. The mmWave spectrum encompasses the frequencies $30-300$ GHz, however the  considered mmWave bands for use in urban cellular systems are at $28$ and $38$ GHz, since it has been shown that rain attenuation and oxygen absorption is low \cite{maccartney2013path,niu2015survey} at those frequencies. The large available bandwidth  will enable data rates of the order of Gigabits-per-second (Gbps), and thus will be able to support future wireless applications such as Vehicle-to-Vehicle (V2V) or Vehicle-to-Infrastructure (V2I) communications. However, to exploit the full potential of mmWave networks for outdoor-to-outdoor wireless communications, a new set of challenges, which are not present in current wireless networks would need to be overcome, such as severe path loss, and sensitivity to blockage and fading \cite{rangan2014millimeter}. 

Various  measurement campaigns \cite{maccartney2013path,park2016millimeter,lee201828,karttunen2017spatially} have revealed that mmWaves incur increased propagation path-loss as compared to the sub-$6$ GHz frequencies used today. One way to compensate for that effect is to employ transmit beamforming \cite{biswas2016performance}, which can increase the probability of the signal arriving successfully at the destination. MmWaves also experience large-scale fading, which, according to measurement campaigns conducted in New York City and Austin, and Texas \cite{maccartney2013path}, can be modeled as normally distributed in the $dB$ domain. 
A similar campaign conducted in Daejeon City, Korea led to similar conclusions \cite{park2016millimeter}.
MmWave wireless communications in densely built cities are also susceptible to blockage by their surrounding environment, e.g., by high-rise buildings, moving vehicles, pedestrians, etc. \cite{zhao201328,molisch2016millimeter,rappaport2017small}.
In such an environment, the  dominant propagation mode of mmWaves is along street canyons by reflecting off of buildings, in a way analogous to ray optics. This is facilitated by the high reflection coefficient of common materials found in cities, such as concrete and glass
\cite{rappaport2013millimeter}.

One way to overcome  blockage effects would be to deploy a dense network of street level mmWave base stations. Although this would  optimize the city wide Line-of-Sight (LoS) coverage rate \cite{palizban2017automation}, it would not provide  Quality-of-Service (QoS) assurances. In fact, a dense network of mmWave base stations can degrade the QoS of users due to interfering base station transmissions \cite{mavromatis2019efficient,wu2018improving}. An alternative solution to overcoming blockage and improve coverage is through relaying \cite{rangan2014millimeter,bai2015coverage,biswas2016performance}, a concept that has been included in the currently applied LTE-Advanced standard \cite{peters2009relay}. 
Multi-hop relaying can improve connectivity \cite{lin2015connectivity} as well as the coverage probability and the transmission capacity of a network \cite{biswas2016performance}. However, a large number of hops requires significant signaling and scheduling overhead to determine the most suitable relays, which in turn can lead to  increased energy consumption and latency. Also, the potential security issues associated with multi-hop relaying cannot be ignored \cite{sun2019secure,jameel2018comprehensive}, as the points at which a malicious user can intercept the transmitted signal increase. On the other hand, $2$-hop relaying in combination with relay beamforming can increase the communication range while avoiding most of the aforementioned issues \cite{soliman2013dual}, at the cost, however, of added complexity for the computation of beamforming weights and phase synchronization \cite{kalogerias2018spatially}. By additionally taking into consideration the 3D structure of the environment, signal blockage can also be avoided by optimal positioning of aerial mobile relays \cite{chenoptimal}. This approach has already been considered for supporting mmWave communication \cite{zhang2019survey}, in the context of Unmanned Aerial Vehicles (UAV) networks. 

\begin{figure}[t]
    \centering
    \includegraphics[scale=0.68]{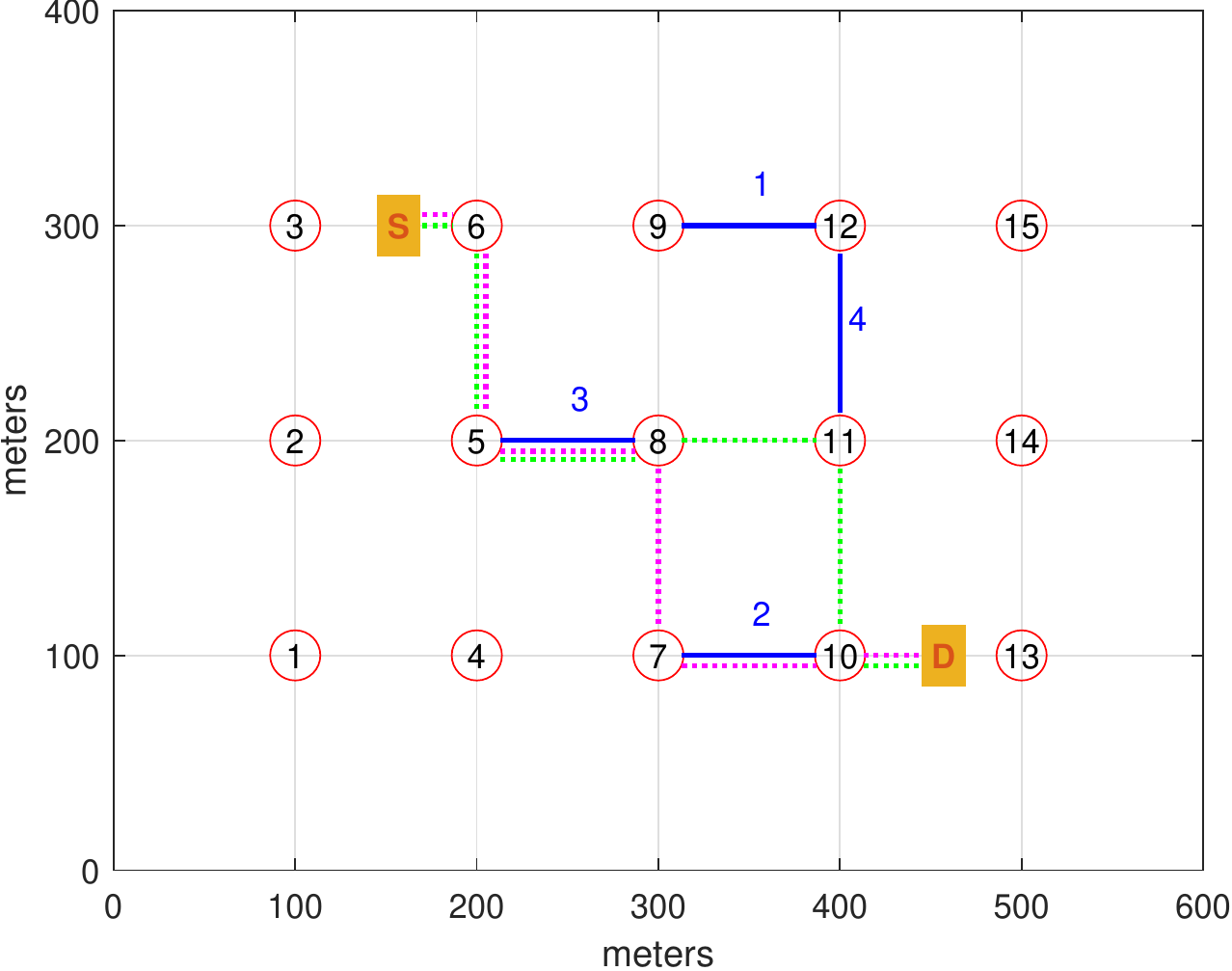}
    \caption{A birds-eye view of the assumed urban topology. Each numbered red circle depicts an intersection, that connects two or more road segments. A solid numbered blue line depicts a relay cluster placed along the respective street segment. The dotted lines depict all mmWave signal propagation paths from the source to the destination, through relay cluster $3$.}
     \label{topology}
\end{figure}

The most widely studied problem in $2$-hop relaying for device-to-device outdoor communication is  relay selection, i.e., determining the minimum number of deployed relays \cite{zheng2015toward},and/or best relay configuration, in order to optimize the quality of communication \cite{wei2016optimal,biswas2016performance,salam2015capacity,turgut2016energy,thi2018proportional}. 
Typically, relay selection requires instantaneous Channel State Information (CSI) between the source and relays, and the relays and destination. When CSI changes with time, optimal relay selection can become a resource-demanding task, as the network needs to \textit{first} compute its channels via the exchange of pilots, and \textit{then} decide on the best relay configuration.  This process takes up bandwidth, wastes power and introduces latency and interference in the network. In the following, any relay selection scheme that conforms to the aforementioned serial procedure will be referred to as \textit{ideal}.

In this paper, we consider point-to-point, relay-assisted mmWave communications in an urban scenario, and propose a new, resource efficient relay selection scheme for overcoming the effects of blockage and severe attenuation, caused by compact urban forms. The proposed scheme is designed to \textit{optimally} enhance QoS in $2$-hop Amplify-and-Forward (AF) cooperative networks and, as compared to the state of the art, induces \textit{significantly lower CSI estimation overhead}, while, providing \textit{substantially better network utilization}.


More specifically, the system model considered consists of static relays deployed in clusters across streets, within a specific area over which the channels have similar statistical characteristics. This is typical in mmWave networks, which are primarily  designed for relatively short distance point-to-point communications.
Assuming a time-slotted system operation, the proposed scheme optimizes QoS in a $2$\textit{-stage fashion}, where, in every time slot, and \textit{simultaneously} with AF beamforming to the destination, each cluster \textit{predictively} selects a new representative ($1$st stage) \textit{to optimally enhance AF beamforming} \textit{at the subsequent time slot} ($2$nd stage). Predictive relay selection is achieved by exploiting channel correlations with current and past networkwide magnitude-only CSI (also known as Received Signal Strength (RSS)) which is \textit{invariant to relay cluster size}, and is measured sequentially during the operation of the system.


In our approach, network QoS is quantified by the \textit{Signal-to-Interference}$+$\textit{Noise Ratio (SINR)} at the destination, which is a standard performance metric, and our goal is to maximize that SINR, subject to a shared power constraint among all relay clusters. Therefore, the optimal beamforming weights need to be computed centrally for all clusters. Nevertheless, the proposed relay selection procedure is conducted in a \textit{completely distributed} manner; each cluster \textit{independently} decides its successor representative for the subsequent time slot by solving a simple local stochastic optimization problem, \textit{without} the need for inter-cluster information exchange.

Exploiting CSI correlations to predictively determine mobile relay movement in cooperative beamforming networks was recently explored in \cite{kalogerias2018spatially}, in a convetional, free-space setting.
Here, we use a similar idea for relay selection. However, the special mmWave signal propagation characteristics make the treatment of the selection problem substantially different than that of \cite{kalogerias2018spatially}. In particular, the diversity resulting from the reflective nature of the mmWave signal propagation, as well as the possibly street-wise varying channel parameters are two features not present in free-space formulations, which induce significant complications herein, as far as both the description of the problem and the development of corresponding efficient implementation techniques are concerned.

   \begin{figure}[t]
       \centering
       \includegraphics[scale = 0.4]{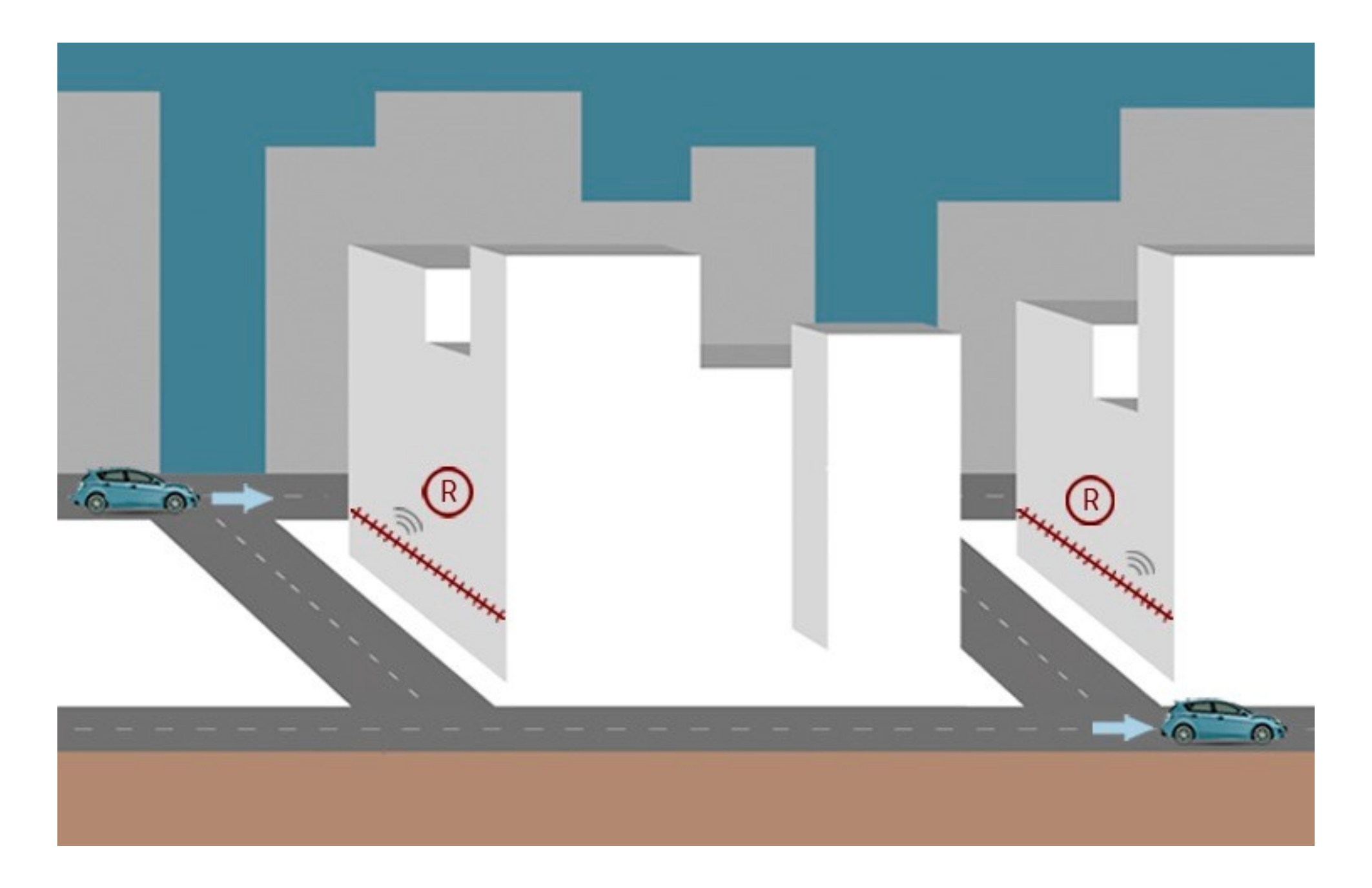}
       \caption{Two relay clusters of our proposed mmWave relay network operating in a city.}
      \label{city}
   \end{figure}

Along with our proposed adaptive relay selection scheme, this paper makes the following additional contributions:

\begin{enumerate}
    \item 
    We propose
    \textit{distributed cooperative} beamforming for SINR maximization in mmWave networks. Our beamforming formulation alone allows for efficient exploitation of the spatial diversity induced by dominant mmWave propagation paths, which is a consequence of the spatial propagation patterns of the mmWave medium \cite{wang2018mmwave,andrews2016modeling,jameel2018propagation}. The CSI induced by the mmWave propagation paths is optimally combined constructively at the destination, resulting in superior network QoS, without the disadvantages of multihop relaying.
  \item As briefly mentioned above, within each time slot, the execution of the proposed relay selection scheme is \textit{completely decoupled} from optimal beamforming. Consequently, optimal communications and optimal relay selection can be performed \textit{in parallel}, resulting in substantially improved time slot utilization. This is due to the predictive nature of the proposed scheme, which allows for the determination of the best cluster representative \textit{before} the start of each time slot. Additionally, we show explicitly that our predictive scheme requires less CSI estimation effort per time slot as compared to the respective ideal scheme, with such reduction being more pronounced as the relay density per cluster increases. This is particularly important in mmWave networks, where dense infrastructure is essential for achieving satisfactory performance \cite{website2}. 
  
  \item We propose a practical and computationally efficient technique for implementing our proposed relay selection scheme. Specifically, the local stochastic problem each cluster is responsible for is replaced by a surrogate based on Sample Average Approximation (SAA) \cite{shapiro2009lectures}, which relies on \textit{predictive Monte Carlo sampling} of the channel uncertainty involved. The proposed technique efficiently exploits spatiotemporal correlations of the mmWave channel structure via a well-designed combination of Kalman filtering and Gaussian process regression, and results in easily computable, near-optimal relay selection policies.
  
   
   %
  
      
   
\end{enumerate}

The effectiveness of the proposed joint beamforming and relay selection system is confirmed through numerical simulations, conducted using synthetically generated CSI data. First, all our numerical results show that the SINR performance of our adaptive relay selection scheme is \textit{almost identical} to the ultimate performance achieved via the respective ideal scheme. On the other extent, our simulations also verify that, as expected, the proposed strategic scheme clearly outperforms any randomized selection policy which does not exploit experience accumulated during system operation. 

Additionally, we examine the effect of different cluster topologies on overall system performance. On the one hand, we confirm that expected network QoS increases with the number of clusters taking part in the communication, also revealing a probably sublinear relevant trend. On the other hand, our simulations corroborate that cluster placement in the city indeed affects the overall QoS of the network users, recognizing the importance of cluster assignment as an open problem for future research.



\section{mmWave Urban Channel Model} \label{channelmodel}
This section is dedicated to the development of a sufficiently detailed urban mmWave channel model, which will be exploited throughout the rest of the paper. Our channel model is versatile enough so that it can be applied to \textit{any} city topology consisting of a densely built area with high-rise buildings, separated by non-curved street canyons. 

To facilitate our presentation, we hereafter consider simplified city topologies such as that of Fig. \ref{topology}, which shows a top view schematic of a particular urban area, where the numbered circles indicate road intersections, and the vertical and horizontal lines connecting those circles denote streets. Due to blockage caused by high-rise buildings, the only way a mmWave signal starting from a source located at $\mathbf{p}_{S}$ can reach its destination at $\mathbf{p}_{D}$ is by  traversing  street segments \cite{wang2018mmwave}. More specifically, the transmitted signal is \textit{spatially diversified} through all sets of consecutive, non-repeating segments from the source to the destination. 
Then, \textit{a (dominant) propagation path} is defined as every such set of traversed street segments whose aggregate length is equal to the \textit{minimum} $\ell_1$\textit{-distance} from the source to the destination. We adopt the convention of \cite{wang2018mmwave}, where the \textit{Line-of-Sight (LoS)} portion of every path is the segment between the transmitting node and  the nearest intersection, while the remaining segments comprise the \textit{Non-Line-of-Sight (NLoS)} portion of the path. All considered paths have common LoS portion, while their NLoS portions differ.


To overcome severe signal attenuation, we deploy clusters of relays across certain street segments, which will beamform the signal to its destination. For simplicity, we assume that each cluster contains evenly spaced relays. At each time instance, only one relay from every cluster, namely, \textit{the cluster representative},  is active. The relays are connected via fiber to a central node via which they can exchange information. A propagation path between the source or destination and any of the cluster representatives, as well as the corresponding LoS and NLoS portions of the path are all defined in exactly the same fashion as in the previous paragraph.

Let $N_c$ be the number of available relay clusters in the network. Also, let $L_r$ be the number of all possible signal paths from $\mathbf{p}_{S}$ to relay cluster $r=1,\ldots,N_c$.
%
The channel between $\mathbf{p}_{S}$ and a relay in cluster  $r$ located at $\mathbf{p}$ is experienced as a combination of all channels across all possible paths between $\mathbf{p}_{S}$ and $\mathbf{p}$. In particular, under the flat fading assumption, the complex channel gain from $\mathbf{p}_{S}$ to point $\mathbf{p}$ along path $i$ can be decomposed as \cite{heath2017introduction}
\begin{align} \label{channel_decomp}
f_{ri}({\mathbf{p}},t)\triangleq\underbrace{f^{PL}_{ri}( \mathbf{p})}_{path-loss}  \underbrace{f^{SH}_{ri} ( \mathbf{p},t)}_{shadowing}  \underbrace{f^{MF}_{ri} ( \mathbf{p},t )}_{multi-path},
\end{align}
where $f^{PL}_{ri}\left( \mathbf{p}\right)$ is the path-loss component, $f^{SH}_{ri} \left( \mathbf{p},t\right)$ the large-scale fading
component (shadowing), and $f^{MF}_{ri} \left( \mathbf{p},t \right)$ the small-scale fading component (multi-path). A similar decomposition holds for the  channel $g_{ri}({\mathbf{p}},t)$ from $\mathbf{p}$ to $\mathbf{p}_D$,  along path $i=1,\ldots,K_r$, where $K_r$ is the number of respective signal paths from cluster $r$ to the destination.

In the mmWave setting, the channel path-loss does not depend on the Euclidean distance between $\mathbf{p}_S$ and $\mathbf{p}$, but rather on their absolute locations (Manhattan distance), and is therefore parametrized separately for each segment \cite{karttunen2017spatially,molisch2016spatially,wang2018mmwave}. 
Let the set of all individually traversed street segments $\tau$ of path $i$ to cluster $r$ be denoted by $\mathcal{S}_{ri}^f$, which includes the segment $\tau_S$ where the source is located, but does not include segment $\tau_r$ where cluster $r$ is located. Similarly, the set of traversed segments of path $i$ from cluster $r$ to the destination, including the segment $\tau_D$ the destination is located but excluding segment $\tau_r$, is $\mathcal{S}_{ri}^g$. In the following, we consider only the source-relay channels $f_{ri}$, for every path $i$ associated with cluster $r$. The discussion for $g_{ri}$ follows in a completely analog manner, and is omitted for brevity.



As in \cite{wang2018mmwave}, we also assume an additional loss $\Delta$ occurring at every intersection, i.e., every propagation path exhibits a total \textit{intersection loss} $\Delta N^f_{r}$, where $N^f_{r}$ are the number of traversed intersections from $\mathbf{p}_S$ to $\mathbf{p}$. Therefore, the overall path-loss component of channel $f_{ri}$ is expressed as
\begin{align} \label{path-loss}
    \hspace{-5pt}f^{PL}_{ri}\hspace{-1pt}\left( \mathbf{p}\right)\hspace{-1.5pt}=\hspace{-1.5pt}10^{-\frac{\Delta N^f_r}{2\cdot 10}} d_{\tau_S}^{-\frac{\alpha_L}{2}} \hspace{-1pt}(d_{\tau_r}^f\hspace{-1pt}(\mathbf{p}))^{-\frac{\alpha_N}{2}} \hspace{-2pt}\prod_{{\tau}\in \mathcal{S}_{ri}^f \setminus \{\tau_S\}}\hspace{-2pt}d_{\tau}^{-\frac{\alpha_N}{2}}\hspace{-0.5pt}, \hspace{-3pt} 
\end{align}
where $d_{\tau_S}$ denotes the length of the LoS segment $\tau_S$, $d_{\tau_r}^f(\mathbf{p})$ is the distance between the intersection of segment $\tau_r$ associated with $f_{ri}$ and location $\mathbf{p}$ in $\tau_r$ (that is, the intersection of $\tau_r$ which is $\ell_1$-closest to the source), and $d_{\tau}$ denotes the length of the $\tau$-th street segment.
We assume that a relay \textit{cannot} be located \textit{exactly on} an intersection, so $d_{\tau_r}^f\neq 0$.

Likewise, the shadowing and multi-path components of the channel experienced across each path may be decomposed on a per-segment basis as 
\begin{align}
f^{SH}_{ri} \big( \mathbf{p},t\big)&= s^f_{\tau_r}(\mathbf{p},t)\prod_{{\tau}\in \mathcal{S}_{ri}^f} s_{\tau}(t)\quad\text{and}  \label{shadowing}\\
 f^{MF}_{ri} \big( \mathbf{p},t\big) &=  q^f_{\tau_r}(\mathbf{p},t) \prod_{\tau \in \mathcal{S}_{ri}^f} q_{\tau}(t)  \label{multi-path},
\end{align} 
where $s_{\tau}$ and $q_{\tau}$ are the shadowing and multi-path terms experienced across segment $\tau$. 

Consequently, by expressing the magnitude of \eqref{channel_decomp} in logarithmic scale ($dB$), we obtain the \textit{additive model}
\begin{align} \label{channeldB}
F_{ri}\left( \mathbf{p},t\right) &\hspace{-1pt}\triangleq\hspace{-1pt} 10\text{log}_{10}\big( \big| f^{PL}_{ri}\big( \mathbf{p}\big) \cdot f^{SH}_{ri} \big( \mathbf{p},t\big) \cdot f^{MF}_{ri} \big( \mathbf{p},t\big) \big|^2 \big)\nonumber \\
&\hspace{-1pt}\triangleq\hspace{-1pt}a^f_{ri}(\mathbf{p})+b^f_{ri}(\mathbf{p},t)+c^f_{ri}(\mathbf{p},t)
\end{align}%
where
\begin{align}
 -a^f_{ri}(\mathbf{p})\hspace{-1pt} &\triangleq \hspace{-1pt}\alpha_L10\text{log}_{10}d_{\tau_S} \hspace{-2pt}+\alpha_N \hspace{-5pt} \sum_{\tau \in \mathcal{S}_{ri}^f} \hspace{-5pt}10\text{log}_{10}d_{\tau} \nonumber\\
 &\quad+a_N10\text{log}_{10}d^f_{\tau_r}(\mathbf{p})+\Delta N^f_{r},  \label{path-lossdB} \\
    b^f_{ri}(\mathbf{p},t) \hspace{-1pt}&\triangleq\hspace{-1pt}\sum_{{\tau}\in \mathcal{S}_{ri}^f} \hspace{-2pt}10\text{log}_{10}|s_{\tau}(t)|^2+ 10\text{log}_{10}|s^f_{\tau_r}(\mathbf{p},t)|^2 \nonumber \\
    &\triangleq  \sum_{{\tau}\in \mathcal{S}_{ri}^f}\hspace{-2pt}\beta_{\tau}(t) +\beta^f_{\tau_r}(\mathbf{p},t)\quad \text{and}  \label{shadowingdB}\\
c^f_{ri}(\mathbf{p},t) \hspace{-1pt}&\triangleq\hspace{-1pt} \sum_{\tau\in \mathcal{S}_{ri}^f}\hspace{-2pt} 10\text{log}_{10}|q_{\tau}(t)|^2 + 10\text{log}_{10}|q^f_{\tau_r}(\mathbf{p},t)|^2 \nonumber \\ 
\textbf{}&\triangleq \hspace{-1pt} \sum_{\tau\in \mathcal{S}_{ri}^f}\hspace{-2pt}\xi_{\tau}(t)+\xi^f_{\tau_r}(\mathbf{p},t). \label{multi-pathdB}
\end{align}
We should note that, in (\ref{shadowingdB}), the \textit{combined} shadowing components pertaining to segments without relays have been separated from the respective term referring to the segment containing the relay cluster. Those terms exhibit distinct statistical behavior, and will be considered separately.

In addition to the above, for every time slot $t$, we assume a phase term $e^{j2\pi \phi_{\tau}(t)}$ for each distinct segment $\tau\in \mathcal{S}_{ri}^f$, $r=1,\ldots,N_c$, $i=1,\ldots,L_r$, where each $\phi_{\tau}(t)$ is uniformly distributed in $[0,1]$. Similarly, for every time slot $t$ and every location $\mathbf{p}$, we assume another phase term $\phi^f_{\tau_r}(\mathbf{p},t)$, also uniformly distributed in $[0,1]$. Across \textit{all} time slots, \textit{all} segments, and \textit{all} locations, \textit{all} phase components are mutually independent, and also independent of the respective channel magnitudes, as well.  %
Then, the channel $f_{ri}(\mathbf{p},t)$ can be reconstructed as 
\begin{align} \label{reconstruct}
f_{ri}(\mathbf{p},t)={e^{\frac{ln(10)F_{ri}(\mathbf{p},t)}{20}}} {e^{j2\pi \Phi^f_{ri}(\mathbf{p},t)}}, 
\end{align}
where $\Phi^f_{ri}(\mathbf{p},t)\triangleq\sum_{\tau\in \mathcal{S}_{ri}^f}\phi_{\tau}(t)+\phi^f_{\tau_r}(\mathbf{p},t)$.

For all segments, we assume a log-normal distribution for modeling  shadowing and multi-path fading \cite{rappaport2017small}. The channel paths $f_{ri}$, $i=1,\ldots, L_r$, are statistically dependent, as they might traverse common segments. Still, it is reasonable to model all shadowing and multi-path components as being mutually independent across different segments, since each segment will exhibit distinct spatial features. 

However, \textit{within} each segment $\tau\in \mathcal{S}_{ri}^f$, $\beta_{\tau}(t)$ is assumed to be zero mean and jointly Gaussian in time, with  correlation between two time slots $k$ and $l$ given by
\begin{align} \label{temporal_corr}
  \mathbb{E}[\beta_{\tau}(k) \beta_{\tau}(l)]\triangleq\eta^2 e^{-|k-l|/\gamma}, 
\end{align}
where $\eta^2$ is the \textit{shadowing power} and $\gamma$ the \textit{correlation time} \cite{kalogerias2018spatially}. Further assuming that the multi-path component $q_\tau(t)$ is white in time with variance $\sigma^2_\xi$ \cite{malmirchegini2012spatial}, the combined log-magnitude terms $z_{\tau}(t)\triangleq\beta_{\tau}(t)+\xi_{\tau}(t),t=1,\ldots,N_T$ are jointly Gaussian with mean zero and covariance 
\begin{align} \label{nonrelay_covar}
\bm{\Sigma}_{\tau} \hspace{-1.5pt}&\triangleq\hspace{-1pt} \eta^2 \hspace{-3.5pt} \begin{bmatrix} 1 & \hspace{-6pt}\ldots & 
e^{-(N_T-1)/\gamma} \\
\vdots & \hspace{-6pt}\ddots & \vdots \\
e^{-(N_T-1)/\gamma} & \hspace{-6pt}\ldots & 1
\end{bmatrix} \hspace{-3pt}+\hspace{-1pt} \sigma^2_{\xi} \mathbf{I}_{N_T} \nonumber \\
\hspace{-1.5pt}&\triangleq\hspace{-1pt} \eta^2 \mathbf{T}+\sigma^2_{\xi} \mathbf{I}_{N_T} \in \mathbb{R}^{N_T \times N_T}.
\end{align}%

\begin{figure}
    \centering
    \includegraphics[scale=0.8]{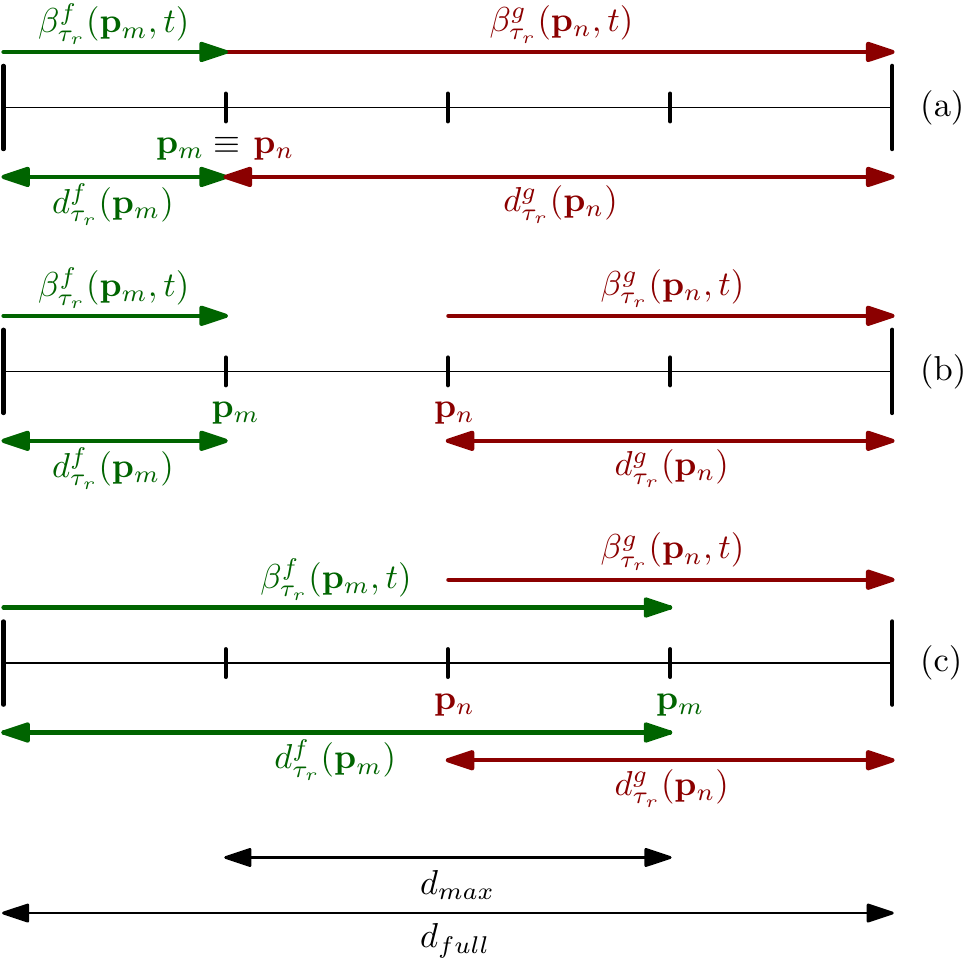}
    \caption{Crosscorrelation structure of the incoming and outgoing channel terms $\beta^f_{\tau_r}(\mathbf{p}_n,t)$, $\beta^g_{\tau_r}(\mathbf{p}_m,t)$ when (a) $d_{\tau_r}^f(\mathbf{p}_m) +d_{\tau_r}^g(\mathbf{p}_n)= d_{full}$, (b) $d_{\tau_r}^f(\mathbf{p}_m) +d_{\tau_r}^g(\mathbf{p}_n) < d_{full}$, and (c)  $d_{\tau_r}^f(\mathbf{p}_m) +d_{\tau_r}^g(\mathbf{p}_n)> d_{full}$, for cluster $r$ with $\delta=3$ relay positions, at a common time slot $t$.}
   \label{fig:Kernel}
\end{figure}

Likewise, the term $\beta^f_{\tau_r}(\mathbf{p},t)$, corresponding to the segment where cluster $r$ is located, is assumed to be jointly Gaussian, \textit{both} in space and time. Specifically, we assume that the individual relays of cluster $r$ can be located at a discrete set of $\delta$ positions across the segment $\tau_r$. At two such positions, say $\mathbf{p}_m$ and $\mathbf{p}_n$, and between any two time slots $k$ and $l$, the spatiotemporal correlation of $\beta_{\tau_r}(\mathbf{p},t)$ is defined as \cite{kalogerias2018spatially}
\begin{align} \label{spatiotemp_corr}
  \mathbb{E}[\beta^f_{\tau_r}(\mathbf{p}_n,k)\beta^f_{\tau_r}(\mathbf{p}_m,l)]\triangleq \mathsf{K}_{FF}(\mathbf{p}_n,\mathbf{p}_m) e^{-|k-l|/ \gamma},
\end{align}%
where the spatial kernel $\mathsf{K}_{FF}$ is given by
\begin{align} \label{spatiotemp_corr2}
  \mathsf{K}_{FF}(\mathbf{p}_n,\mathbf{p}_m)\triangleq \eta^2 e^{-\hspace{1pt}\norm{\mathbf{p}_n-\mathbf{p}_{m}}_2/{\beta}}.
\end{align}%
We further assume that the "incoming" and "outgoing" shadowing terms $\beta^f_{\tau_r}(\mathbf{p},t)$ and $\beta^g_{\tau_r}(\mathbf{p},t)$ at positions $\mathbf{p}_m$ and $\mathbf{p}_n$ and between time slots $k$ and $l$ are \textit{themselves} correlated as 
\begin{align}
  \label{kernel}
  \mathbb{E}[\beta^f_{\tau_r}(\mathbf{p}_n,k)\beta^g_{\tau_r}(\mathbf{p}_m,l)]\triangleq
    \mathsf{K}_{FG}(\mathbf{p}_n,\mathbf{p}_m) 
        e^{-|k-l|/ \gamma},
\end{align}
where the \textit{crosscorrelation} kernel $\mathsf{K}_{FG}$ is defined as
\begin{align}
  \label{kernel2}
    \mathsf{K}_{FG}(\mathbf{p}_n,\mathbf{p}_m)
    \triangleq
        \eta^2 e^{({ \epsilon \hspace{1pt} \norm{\mathbf{p}_n-\mathbf{p}_m}_2-d_{max}}) /{\beta}},
\end{align}
with $\epsilon =1$ for $d_{\tau_r}^f(\mathbf{p}_m) +d_{\tau_r}^g(\mathbf{p}_n)\geq d_{full}$ and $\epsilon=-1$ otherwise,
and where $d_{full}$ is the length of segment $\tau_r$, and $d_{max}$ is the furthest possible distance between two discrete relay positions of a cluster. This kernel describes the correlation between the incoming and outgoing channels at each cluster, at different locations and at different time slots. Intuitively, correlation should be proportional to the size of the part of the segment which is traversed by both channels, if such a part exists (see Fig. \ref{fig:Kernel}c). Otherwise, as the distance between the locations where the two channels are respectively experienced increases, their correlation should be decreasing (see Fig. \ref{fig:Kernel}b). The proposed kernel captures precisely the behavior outlined above, while resulting in a valid cross-covariance structure for $\beta_{\tau_r}^f$ and $\beta_{\tau_r}^g$.

As above, assuming that $q^f_{\tau_r}(\mathbf{p},t)$ and $q^g_{\tau_r}(\mathbf{p},t)$ are both white in \textit{both} space and time, as well as mutually independent, the collection of combined terms
\begin{align}
  \begin{bmatrix}
  z^f_{\tau_r}(\mathbf{p}_i,t) \\
  z^g_{\tau_r}(\mathbf{p}_i,t)
  \end{bmatrix}\hspace{-3pt}\triangleq\hspace{-3pt}
  \begin{bmatrix}
  \beta^f_{\tau_r}(\mathbf{p}_i,t)+\xi^f_{\tau_r}(\mathbf{p}_i,t)\\
  \beta^g_{\tau_r}(\mathbf{p}_i,t)+\xi^g_{\tau_r}(\mathbf{p}_i,t)
  \end{bmatrix},
\end{align}
for $i=1,\ldots,\delta$ and $t=1,\ldots,N_T$, are Gaussian with mean zero and covariance $\bm{\Sigma}_{\tau_r}\in \mathbb{R}^{2 \delta N_T \times 2 \delta N_T}$ given by
\begin{align}\label{relay_covar}
\bm{\Sigma}_{\tau_r}\triangleq \mathbf{T}\otimes \mathbf{K}+\sigma^2_{\xi} \mathbf{I}_{2 \delta N_T},
\end{align}
where $\otimes$ indicates Kronecker product, and the per-slot cross-covariance matrix $\mathbf{K}\in \mathbb{R}^{2 \delta \times 2 \delta}$ is defined as
\begin{align}\label{Kmatrix}
    \mathbf{K}\triangleq\begin{bmatrix}\mathsf{K}_{FF} & \mathsf{K}_{FG}\\ \mathsf{K}_{FG} & \mathsf{K}_{GG} \end{bmatrix},
\end{align}
where, \textit{overloading notation}, $\mathsf{K}_{FF}$, $\mathsf{K}_{GG}$ and $\mathsf{K}_{FG}$ are correlation matrices corresponding to the kernels \eqref{spatiotemp_corr2} and \eqref{kernel2}, respectively, each evaluated on all $\delta^2$ pairs of possible positions across segment $\tau_r$, according to some common order.

\section{Joint Beamforming and Relay Selection}\label{scheme}

Determining the cluster topology,  i.e.,  the number of deployed clusters as well as their locations  is an important problem that will be studied in our future work, and could draw from current literature on optimal relay placement  \cite{zheng2015toward,ghatak2019relay,ding2018hop}.
Our proposed  scheme operates assuming that the clusters change at a low rate, and  focuses on the time period over which the clusters have been optimally determined and are fixed. During that time, the  statistical model of the channel stays the same; however, the channel itself changes.

In every time slot, the proposed system \textit{jointly} performs  \textit{beaforming} and \textit{relay selection}, by addressing a $2$-stage stochastic problem \cite{kalogerias2018spatially}.
Before going into the details (and the advantages) of each stage separately, we should note that although the $2$-stage problem refers to the necessary actions needed to be taken during a single time slot, in practice these actions refer to two consecutive time slots, due to the availability of the required CSI. More specifically, during time slot $t$, \textit{both} current beamforming weights of the cluster representative are calculated (corresponding, as discussed below, to the $2nd$ stage problem at time slot $t$), \textit{and} the relays from all clusters \textit{to be selected at the next time slot} are determined (which corresponds to the $1st$ stage problem at time slot $t+1$). Both tasks (beamforming and relay selection) are based on current CSI, as well as past CSI of cluster representatives selected up to time slot $t$. 

We assume that $2$-hop relaying is used to assist the communication between $\mathbf{p}_{S}$ and $\mathbf{p}_{D}$. The whole network is assumed to operate for $N_T$ time slots. In each time slot $t=1,\ldots,N_T$, the source at $\mathbf{p}_S$ transmits the signal $\sqrt{P_S}s(t)$, where $s(t)$ is an information symbol with $\text{E}[|s(t)|^2]=1$, and $P_S > 0$ the source transmission power. The signal received at the representative relay of \textit{each} cluster $r$, located at $\mathbf{p}_r(t)$ is,
\begin{align}\label{relay_signal}
R_r(t)=\sum_{i=1}^{L_r} \sqrt{P_S}f_{ri}(t) s(t)+n_{r}(t),
\end{align}
where $n_{r}(t)\sim \mathcal{CN}(0,\sigma^2)$ is the reception noise at cluster $r$. Working in an Amplify-and-Forward (AF) fashion, each cluster representative modulates its received signal $R_r(t)$ by a complex weight $w_r(t)$ and re-transmits it. Note that a mmWave signal arriving at $\mathbf{p}_D$ directly from the source and without the help of a relay has negligible power, and can be ignored. Therefore, the aggregate signal received at the destination from all relay representatives is
\begin{align} \label{signal_dest}
  y_D(t) &=\sum_{r=1}^{N_c} \sum_{k=1}^{K_r} w_r(t)g_{rk}(t)R_r(t)+n_D(t) \nonumber \\
  &=\underbrace{\sqrt{P_S}\sum_{r=1}^{N_c} w_r(t) \sum_{k=1}^{K_r} \sum_{i=1}^{L_r} g_{rk}(t)f_{ri}(t)s(t)}_{signal} \nonumber \\&\quad\quad + \underbrace{\sum_{r=1}^{N_c} w_r(t)\sum_{k=1}^{K_r}  g_{rk}(t)n_r(t)+n_D(t)}_{interference \,+\, destination\, noise},
\end{align}%
where $n_D(t) \sim \mathcal{CN}(0,\sigma_D^2)$ is the reception noise at $\mathbf{p}_D$. 

\subsection{Optimal Beamforming for $2$-hop relaying} 
We extend the distributed relay beamforming schemes studied in \cite{havary2008distributed,zheng2009collaborative}, to the significantly more complex setting of urban mmWave relay networks. Here, distributed beamforming is considered for enforcing relay cluster cooperation, such that \textit{all individual signal paths} forwarded from \textit{all} relay clusters are combined constructively at the destination.

At every time slot $t$, the goal is to obtain the respective beamforming weights to be used by each cluster, $\mathbf{w}(t)\triangleq[w^*_1(t), \ldots, w^*_{N_c}(t)]^{\text{T}}\in \mathbb{C}^{N_c\times 1}$, such that the SINR at $\mathbf{p}_D$ is maximized, subject to a total transmission power budget $P_C>0$ over all relay clusters. Define the vectors 
\begin{align}
    \mathbf{f}_r(\mathbf{p},t) &\triangleq [f_{r1}(\mathbf{p},t),\ldots, f_{rL_r}(\mathbf{p},t)]^{\text{T}} \in \mathbb{C}^{L_r \times 1} , \label{fvector}\\
    \mathbf{g}_r(\mathbf{p},t) &\triangleq [g_{r1}(\mathbf{p},t),\ldots, g_{rK_r}(\mathbf{p},t)]^{\text{T}} \in \mathbb{C}^{K_r \times 1}  \label{gvector},
\end{align}%
$r=1,\ldots, N_c$. Then, after dropping dependence on $(t)$ and $(\mathbf{p}_r(t),t)$ for brevity, the SINR is maximized by solving \cite{havary2008distributed} 
\begin{align} \label{optimization}
\begin{split}
    \underset{\mathbf{w}}{\text{maximize}} &\quad \frac{\mathbf{w}^{\text{H}}\mathbf{R}\mathbf{w}}{\mathbf{w}^{\text{H}}\mathbf{Q}\mathbf{w}+\sigma_D^2} \\
    \text{subject to} &\quad  \mathbf{w}^{\text{H}}\mathbf{D}\mathbf{w} \leq P_C 
\end{split} \, ,
\end{align}
where
\begin{align} 
\mathbf{R} &\triangleq P_S \mathbf{hh}^{\text{H}}, \,\, \mathbf{h}\triangleq[\mathbf{1}^{\text{T}} \mathbf{g}_1 \mathbf{1}^{\text{T}}\mathbf{f}_1,\ldots,\mathbf{1}^{\text{T}} \mathbf{g}_{N_c} \mathbf{1}^{\text{T}} \mathbf{f}_{N_c}]^{\text{T}},  \\
\mathbf{D} &\triangleq P_S \text{diag}\big(|\mathbf{1}^{\text{T}}\mathbf{f}_1|^2,\ldots, |\mathbf{1}^{\text{T}}\mathbf{f}_{N_c}|^2 \big)+ \sigma^2 \mathbf{I}_{N_c} \,\,\, \text{and} \\
\mathbf{Q} &\triangleq\sigma^2 \text{diag}\big(|\mathbf{1}^{\text{T}}\mathbf{g}_1 |^2,\ldots,|\mathbf{1}^{\text{T}}\mathbf{g}_{N_c} |^2\big).
\end{align}
and $\mathbf{1}$ is the all-ones vector. 
A crucial technical property of problem \eqref{optimization} is that its optimal can be explicitly expressed as \cite{havary2008distributed, kalogerias2018spatially} 
\begin{align}\label{optimalvalue_expand}
  V(t) &=\sum_{r=1}^{N_c} \frac{P_CP_S |\mathbf{1}^{\text{T}}\mathbf{f}_r|^2 |\mathbf{1}^{\text{T}}\mathbf{g}_r |^2}{P_S\sigma_D^2 |\mathbf{1}^{\text{T}}\mathbf{f}_r|^2+P_C \sigma^2|\mathbf{1}^{\text{T}}\mathbf{g}_r |^2+\sigma^2\sigma_D^2} \nonumber \\ 
  &=\sum_{r=1}^{N_c}V_I\big(\boldsymbol{S}_r(\mathbf{p}_r(t),t)\hspace{-0.5pt}\big),
\end{align}%
where $\boldsymbol{S}_r\left(\mathbf{p}_r(t),t\right)$ is a vector of all random variables referring to the shadowing, multi-path, and phase terms of all \textit{unique} segments traversed for all paths from $\mathbf{p}_S$ to $\mathbf{p}_D$, which also pass through each cluster representative, located at $\mathbf{p}_r(t)$. As can be seen from  \eqref{optimalvalue_expand}, $V(t)$ depends on the relay positions at time slot $t$. Thus, by optimally positioning the relays, $V(t)$ can be further maximized. This problem is explored in the next subsection.

Interestingly enough, it turns out that the optimal beamforming vector that achieves \eqref{optimalvalue_expand} enjoys an explicit form similar to that used in the free space scenario \cite{havary2008distributed}, i.e.,
\begin{align}\label{optimalweights}
    \mathbf{w}_{opt}(t)= \sqrt{P_C}\mathbf{D}^{-\frac{1}{2}} \dfrac{\mathbf{\text{v}}_{max}}{\Vert \mathbf{\text{v}}_{max} {\Vert}_2},
  \end{align}
where the \textit{alignment vector} $\mathbf{\text{v}}_{max} \in \mathbb{C}^{N_c\times 1}$ is defined as
\begin{align}\label{optimalweights_expand}
    \hspace{-2pt}\mathbf{\text{v}}_{max}\hspace{-1.5pt}\triangleq\hspace{-4pt} \begin{bmatrix}
\dfrac{P_S\mathbf{1}^{\text{T}} \mathbf{g}^*_1 \mathbf{1}^{\text{T}}\mathbf{f}^*_1}{P_S\sigma_D^2 |\mathbf{1}^{\text{T}}\mathbf{f}_1|^2+P_C \sigma^2|\mathbf{1}^{\text{T}}\mathbf{g}_1 |^2+\sigma^2\sigma_D^2} \\
\vdots \\
\dfrac{P_S\mathbf{1}^{\text{T}} \mathbf{g}^*_{N_c} \mathbf{1}^{\text{T}}\mathbf{f}^*_{N_c}}{P_S\sigma_D^2 |\mathbf{1}^{\text{T}}\mathbf{f}_{N_c}|^2+P_C \sigma^2|\mathbf{1}^{\text{T}}\mathbf{g}_{N_c} |^2+\sigma^2\sigma_D^2}
\end{bmatrix}\hspace{-2pt},\hspace{-2pt}
\end{align}
and where $|\mathbf{1}^{\text{T}}\mathbf{f}_r|^2$ and $|\mathbf{1}^{\text{T}}\mathbf{g}_r|^2$ are the incoming and outgoing aggregate channels at $\mathbf{p}_r$, respectively. In other words, each cluster representative at $\mathbf{p}_r$ does \textit{not} need to estimate the individual channels from every propagation path, but rather only the aggregate channel from all propagation paths. In practice, this can be computed by the selected relay via the exchange of pilots.

 One may also observe that, while the $i-$th element of ${\mathbf v}_{max}$ can be estimated by the $i$-th relay only,  the vector norm in \eqref{optimalweights} involves the source and destination channels of all cluster representatives who will beamform at the current time. Therefore, that scalar  will have to be computed centrally and then distributed to all clusters. This can be done through a high-speed, optical fiber based, backhaul network, that connects all relay clusters, as well as all relays within a cluster, with each other \cite{website1,website1}. Clearly, for the source and destination which are, e.g. moving vehicles, no wired connection to the backhaul exists. The beamforming stage requires $\mathcal{O}(N_c)$ operations. 

As a final remark, we should also note that in practice, during this beamforming step, phase synchronization is required to take care of local oscillator phase offsets. Distributed beamforming synchronization is an active field of research  \cite{mudumbai2009distributed,nasir2016timing}, that has also been studied in the context of mmWaves \cite{bai2018cooperative,koschel2012frequency,ulusoy201360}. Here, we assume that the backhaul network can also take care of the synchronization between the relay clusters.

\subsection{Optimal Relay Selection for $2$-hop relaying}
 

At every time slot, each cluster must decide which is the appropriate relay to be used for beamforming. Typically, this would \textit{first} require estimating the respective channel of every relay in the cluster, and \textit{then} deciding upon the strongest one. Clearly, this decision making procedure not only wastes power and bandwidth during CSI estimation, but also induces extra delay \textit{before} optimized communication can take place \textit{within} each time slot. This delay is significant, especially if the number of relays per cluster is large.
In this subsection, we propose a new scheme for adaptive relay selection \textit{which completely avoids this overhead}, thus resulting in much better time slot utilization. 


More specifically, the proposed relay selection scheme is based on \textit{transferring} the implementation of the relay selection procedure from the current time slot, to the previous time slot. In other words, relay selection would be implemented \textit{predictively} by efficiently exploiting the statistical model of the mmWave channel, \textit{before} the respective time slot starts. This immediately results in the complete elimination of the "waiting delay" discussed above; indeed, if predictive relay selection is sufficiently accurate, then the cluster representatives at each time slot can be \textit{predetermined}, \textit{before} the slot starts. This means that relay selection and beamforming can be completely decoupled within each time slot, and thus can be parallelized. What is more, as a desirable byproduct of eliminating this "waiting delay", the proposed scheme also comes with a substantial reduction on the amount of CSI required for relay selection, as well as significant power savings. See Section \ref{timeslotphase} for a more detailed discussion. 

We now describe the proposed relay selection scheme in detail. As described above, at time slot $t$, we are interested in deciding on the best relay representatives from \textit{all} clusters to participate in beamforming at time slot $t+1$, such that the networkwide SINR, $V(t+1)$, is maximized.  However, at the current time slot $t$, future CSI needed for evaluating $V(t+1)$ is \textit{not} yet available. Nevertheless, a reasonable \textit{causal} criterion for optimal relay selection is to maximize a \textit{projection} of $V(t+1)$ on information available at time slot $t$. Following this path, we propose to maximize an \textit{MMSE predictor} of $V(t+1)$ relative to the collection $\mathcal{C}_r(t)$ of all \textit{magnitude} CSI, or RSS, from the segments of all propagation paths associated with \textit{all} previously selected representatives of cluster $r$, as well as the positions of the representatives themselves, up until and including $t$. Then, due to the additive structure of \eqref{optimalvalue_expand}, each cluster $r$ can \textit{independently} solve
\begin{align} \label{conditional}
   \begin{split}
    \underset{\mathbf{p}}{\text{maximize}} &\quad  \mathbb{E}\left[V_I(\boldsymbol{S}_r(\mathbf{p},t+1))|\hspace{0.5pt} \mathcal{C}_r(t) \right]\\
    \text{subject to} &\quad \mathbf{p}\in \mathscr{C}_r(t)
    \end{split},
\end{align}%
where $\mathscr{C}_r\left(\cdot\right)$ constitutes the set of candidate relays within the cluster which can potentially be selected. This set can either be \textit{unconstrained}, including any relay within the cluster, or \textit{constrained} to only a subset of relays within the cluster.
Next, define the sets $\mathcal{S}_r^f=\cup_{i=1}^{L_r} \mathcal{S}_{ri}^f$ and $\mathcal{S}_r^g=\cup_{i=1}^{K_r} \mathcal{S}_{ri}^g$. Then, at every feasible location $\mathbf{p}\in\mathscr{C}_r(t)$, the objective of \eqref{conditional} may be expressed as
\begin{align} \label{objective}
\begin{split}
    &\hspace{-10pt}\mathbb{E}[V_I(\boldsymbol{S}_r(\mathbf{p},t+1))\,|\, \mathcal{C}_r(t)] \\ &=\int  V_I (\mathbf{p}, \boldsymbol{s})   p_{\boldsymbol{S}_r(\mathbf{p},t+1)| \mathcal{C}_r(t)}(\boldsymbol{s}) \text{d}\boldsymbol{s},    
\end{split}
\end{align}
where, dropping dependence on $(\mathbf{p},t+1)$, $V_I$ may be reexpressed in a more integration-friendly form as
\begin{align} \label{theVI}
\begin{split}
   &V_I (\cdot,\mathcal{Z}_r^f, \varphi^f,\mathcal{Z}_r^g, \varphi^g) \\ &= \frac{P_CP_S\mathsf{F}\big(\mathcal{Z}_r^f, \varphi^f \big) \mathsf{G}\big(\mathcal{Z}_r^g , \varphi^g \big) }{P_S\sigma_D^2\mathsf{F}\big( \mathcal{Z}_r^f , \varphi^f \big)+P_C \sigma^2 \mathsf{G}\big( \mathcal{Z}_r^g , \varphi^g\big)+\sigma^2\sigma_D^2},
\end{split}
 \end{align}
with $\mathsf{F}$ being a function of the sets $\mathcal{Z}_r^f=\{z^f_{\tau_r},\{z_\tau \}_{\tau \in \mathcal{S}_r^f}\}$, and $\varphi^f_r=\{\phi^f_{\tau_r},\{\phi_\tau\}_{\tau \in \mathcal{S}_r^f}\}$, corresponding to the combined shadowing plus multi-path, and phase terms of the unique segments traversed in all paths between the source and cluster $r$, and respectively for $\mathsf{G}$, $\mathcal{Z}_r^g$ and $\varphi_r^g$. Analytical expressions for $\mathsf{F}$ and $\mathsf{G}$ are presented in Appendix \ref{appendix:b}.

By a slightly tedious but straightforward procedure, it may be shown that the joint conditional density of all random variables contained in vector $\boldsymbol{S}_r(\mathbf{p},t+1)$ relative to $\mathcal{C}_r(t)$ can be expressed as (by overloading notation)
\begin{align}
    &\hspace{-14pt}p_{\boldsymbol{S}_r(\mathbf{p},t+1)|\mathcal{C}_r(t)}(\mathcal{Z}_r^f, \varphi^f,\mathcal{Z}_r^g, \varphi^g)\notag\\ &\hspace{-9pt}=
     \mathcal{N}([z_{\tau_r}^f\,z_{\tau_r}^g];\bm{\mu}^{t+1|t}_{\tau_r}(\mathbf{p}),\bm{\Sigma}_{\tau_r}^{t+1|t}(\mathbf{p}))\notag\\ &\times \mathcal{U}({\phi}_{\tau_r}^f; 0,1 )\hspace{1pt}\mathcal{U}({\phi}_{\tau_r}^g; 0,1 )\notag\\
    &\quad\times \hspace{-1pt} \prod_{{\tau}\in \mathcal{S}^f_r \cup \mathcal{S}^g_r } \hspace{-1pt}\mathcal{N}({z}_{\tau}; \mu_{\tau}^{t+1|t},(\sigma_{\tau}^{t+1|t})^2)  \hspace{1pt}\mathcal{U}({\phi}_{\tau}; 0,1 ),\hspace{-13pt} \label{posterior_1}
\end{align}
where $\mathcal{U}(\cdot;0,1)$ denotes the uniform density on $[0,1]$, and where $\bm{\mu}^{t+1|t}_{\tau_r}(\mathbf{p})$, $\bm{\Sigma}_{\tau_r}^{t+1|t}(\mathbf{p})$, $\mu_{\tau}^{t+1|t}$ and $(\sigma_{\tau}^{t+1|t})^2$ constitute the corresponding posterior statistics.

Note that, although phase information at time slot $t+1$ is present in  \eqref{theVI}, the objective \eqref{objective} is independent of phase information \textit{at past time slots}. This is because of the standard assumption that, for each segment, the phase component of the channel is white in time and space, and mutually independent of the respective phase component of all other segments. Indeed, one may readily observe that, in \eqref{posterior_1}, all distributions associated with channel phases are uniform in $[0,1]$, which is precisely the prior distribution of all phase components, for all segments taking part in the communication.

From the discussion above, it follows that tractably evaluating \eqref{objective} is a challenging task. As it might be expected, the first step towards evaluation of \eqref{objective} is the efficient determination of the aforementioned predictors. This is the subject of the next two subsections.

\subsubsection{Channel Prediction for Cluster-free Segments}
The shadowing component of the channel for a cluster-free segment $\tau$, $\beta_{\tau}$, is a Gaussian process evolving in time, which may also be represented as a \textit{stable autoregression of order} $1$. Indeed, it may be easily shown that, at every segment $\tau\in\cup_{r=1}^{N_c}(\mathcal{S}_r^f \cup \mathcal{S}_r^g)$, $\beta_{\tau}$ can be represented via the stochastic difference equation \cite{kalogerias2017spatially}
\begin{align}
\label{autoreg1}
    \beta_{\tau}(t)&=\kappa \beta_{\tau}(t-1)+w_{\tau}(t),\quad t=1,\ldots,N_T, 
\end{align}
where  $\kappa\triangleq e^{-1/{\gamma}}$, $\beta_{\tau}(0)\hspace{-1pt} \sim\hspace{-1pt} \mathcal{N}(0,\eta^2)$, with the latter being independent of $w_{\tau}(t)\hspace{-1pt}\overset{i.i.d.}{\sim}\hspace{-1pt} \mathcal{N}(0,(1-\kappa^2)\eta^2)$, $t=1,\ldots,N_T$.

At the same time, across time slots, the shadowing process $\beta_{\tau}(t)$ cannot be measured directly. Instead, it may be considered as \textit{corrupted by unpredictable noise}, due to the presence of the multi-path component $\xi_{\tau}(t)$; indeed, at each time slot $t$ and segment $\tau$, the term $z_{\tau}(t)=\beta_{\tau}(t)+\xi_{\tau}(t)$ is observed.

Now, for every segment $\tau\in \cup_{r=1}^{N_c}(\mathcal{S}_r^f \cup \mathcal{S}_r^g)$, define the vector
\begin{align} \label{observations_tau}
\mathbf{m}_{\tau}^{1:t}&\triangleq[z_{\tau}(1), \ldots, z_{\tau}(t) ]^\text{T} \in \mathbb{R}^{t \times 1}, 
\end{align}
which contains all observable CSI magnitudes associated with that segment, up to time $t$.
Then, exploiting the autoregressive representation of \eqref{autoreg1}, it follows that the posterior distribution of $z_{\tau}(t+1)$ relative to $\mathbf{m}_{\tau}^{1:t}$ 
is Gaussian with conditional mean and variance given by
\begin{align}
\mu_{\tau}^{t+1|t}&=\kappa\beta_{\tau}^{t|t}\quad\text{and} \label{condmean_tau1}\\
(\sigma_{\tau}^{t+1|t})^2&=\kappa^2\rho^{t|t}_{\beta_{\tau}}+(1-\kappa^2)\eta^2+\sigma_{\xi}^{2}, \label{condsigma_tau2}
\end{align}
respectively, where, by definition,
\begin{align}
\beta_{\tau}^{t|t}&\triangleq \mathbb{E}[\beta{_\tau}(t)|\mathbf{m}_{\tau}^{1:t}]\quad\text{and}\\
\rho^{t|t}_{\beta_{\tau}}&\triangleq\mathbb{E}[(\beta{_\tau}(t)-\mathbb{E}[\beta{_\tau}(t)|\mathbf{m}_{\tau}^{1:t}])^2|\mathbf{m}_{\tau}^{1:t}]
\end{align} are the conditional mean and variance of $\beta_{\tau}(t)$ relative to $\mathbf{m}_{\tau}^{1:t}$, respectively. Therefore, determination of $\mu_{\tau}^{t+1|t}$ and $(\sigma_{\tau}^{t+1|t})^2$ is \textit{equivalent to} that of $\beta_{\tau}^{t|t}$ and $\rho^{t|t}_{\beta_{\tau}}$, respectively, for all $t=1,\ldots,N_T$. Again due to the autoregressive structure of \eqref{autoreg1}, the latter pair of conditional estimates may be evaluated \textit{recursively} via a Kalman filter, achieving constant computational complexity per time slot. Specifically, for every $t=1,\ldots,N_T$, both  $\beta_{\tau}^{t|t}$ and $\rho^{t|t}_{\beta_{\tau}}$ may be evaluated recursively via the updates
\begin{align}
\beta_{\tau}^{t|t}&=\kappa\beta_{\tau}^{t-1|t-1} + K^t (z_{\tau}(t) - \kappa\beta_{\tau}^{t-1|t-1}),\label{Kalman1}\\
\rho^{t|t}_{\beta_{\tau}}&=
K^t \sigma_\xi^2\quad \text{and}\label{Kalman2}\\
K^t &= \dfrac{\kappa^2 \rho^{t-1|t-1}_{\beta_{\tau}}+(1-\kappa^2)\eta^2}{\kappa^2 \rho^{t-1|t-1}_{\beta_{\tau}}+(1-\kappa^2)\eta^2+\sigma_\xi^2},\label{Kalman3}
\end{align}
initialized by setting $\beta_{\tau}^{0|0}=0$ and $\rho^{0|0}_{\beta_{\tau}}=\eta^2$, stemming from the statistics of the initial condition $\beta_{\tau}(0)$. By direct comparison of \eqref{condmean_tau1} and \eqref{condsigma_tau2} to the Kalman filter equations \eqref{Kalman1}, \eqref{Kalman2} and \eqref{Kalman3}, it is easy to derive an algorithm for the \textit{direct} recursive evaluation of $\mu_{\tau}^{t+1|t}$ and $(\sigma_{\tau}^{t+1|t})^2$, comprised, for $t=1,\ldots,N_T$, by the dynamic equations
\begin{align}
\mu_{\tau}^{t+1|t}&=\kappa(1-K^t)\mu_{\tau}^{t|t-1} + \kappa K^t z_{\tau}(t),\\
(\sigma_{\tau}^{t+1|t})^2 &= (1+\kappa^2 K^t)  \sigma_\xi^2 +(1-\kappa^2)\eta^2\quad \text{and}\\
K^t &= \dfrac{(\sigma_{\tau}^{t|t-1})^2-\sigma_\xi^2}{(\sigma_{\tau}^{t|t-1})^2},
\end{align}
initialized by setting $\mu_{\tau}^{1|0}=0$ and $(\sigma_{\tau}^{1|0})^2=\eta^2+\sigma_\xi^2$.

For each cluster-less segment $\tau\in \cup_{r=1}^{N_c}(\mathcal{S}_r^f \cup \mathcal{S}_r^g)$, the corresponding Kalman filter may be implemented either centrally within each cluster, or in a completely distributed fashion, where each cluster-less segment is responsible for tracking its own channel, and then for distributing its estimate to the associated cluster, responsible for the actual relay selection.

\subsubsection{Channel Prediction for Segments Containing Clusters}

Next, consider the segment $\tau_r$, containing cluster $r$. Then, if we define $\bm{z}_{\tau_r}^{f,g}\triangleq[z_{\tau_r}^f\,z_{\tau_r}^g]^{\text{T}}$ and store all CSI measurements of every previously selected representative of cluster $r$ in 
\begin{align}
   \hspace{-3pt}  \mathbf{m}_{\tau_r}^{1:t}&= 
  [ \bm{z}_{\tau_r}^{f,g}(\mathbf{p}_r(1),1), \ldots,\bm{z}_{\tau_r}^{f,g}(\mathbf{p}_r(t),t)]^{\text{T}} \in \mathbb{R}^{2t \times 1},
\end{align}
then, for each location $\mathbf{p}\in \mathscr{C}_r(t)$, the mean vector and covariance matrix of the Gaussian random vector $\bm{z}_{\tau_r}^{f,g}(\mathbf{p},t+1)$ conditioned on $\mathbf{m}_{\tau_r}^{1:t}$ are
\begin{align} 
  \hspace{-10pt}\bm{\mu}_{\tau_r}^{t+1|t}(\mathbf{p}) &\hspace{-2pt}=\hspace{-2pt}(\bm{\bar{\sigma}}_{\tau_r}^{1:t}(\mathbf{p}))^\text{T}(\bm{\bar{\Sigma}}^{1:t}_{\tau_r})^{-1}\mathbf{m}_{\tau_r}^{1:t} \in \mathbb{R}^{2 \times 1} \label{condmean_taur}\\
 \hspace{-10pt}\bm{\Sigma}_{\tau_r}^{t+1|t}(\mathbf{p}) &\hspace{-2pt}=\hspace{-2pt} \bar{\mathbf{K}}\hspace{-1pt}-\hspace{-1pt}(\bm{\bar{\sigma}}_{\tau_r}^{1:t}(\mathbf{p}))^\text{T}(\bm{\bar{\Sigma}}^{1:t}_{\tau_r})^{-1}\bm{\bar{\sigma}}_{\tau_r}^{1:t}(\mathbf{p}) \in \mathbb{R}^{2 \times 2},\hspace{-8pt} \label{condSigma_taur}
\end{align}
respectively, where 
\begin{align}\label{Kernelbar}
\bar{\mathbf{K}}=\begin{bmatrix}
\eta^2+\sigma^2_{\xi}& \eta^2 e^{-d_{max}/\beta}\\
\eta^2 e^{-d_{max}/\beta}&\eta^2+\sigma^2_{\xi}\\
\end{bmatrix},   
\end{align}
and  $\bm{\bar{\Sigma}}^{1:t}_{\tau_r} \in \mathbb{R}^{2t\times 2t}$, $\bm{\bar{\sigma}}_{\tau_r}^{1:t} \in \mathbb{R}^{2t \times 2}$ are sampled for every time slot until $t$ from $\bm{\Sigma}_{\tau_r}\in \mathbb{R}^{2 \delta N_T \times 2 \delta N_T}$, at the positions that correspond to the distance between the candidate location $\mathbf{p}$ and the respective locations where the incoming and outgoing channels of segment $\tau_r$ have been experienced \textit{so far}.
We should note that unlike before, \eqref{condmean_taur} and \eqref{condSigma_taur} cannot be estimated using a Kalman filter, but rather, using full-blown Gaussian process regression. The dominant operation of \eqref{condmean_taur} and \eqref{condSigma_taur} is the inversion of the covariance matrix $\bm{\bar{\Sigma}}^{1:t}_{\tau_r}$. The computational complexity of this inversion is of the order of $\mathcal{O}(t^3)$ operations, and grows with time due to conditioning on past CSI. Nevertheless, the complexity can be reduced to $\mathcal{O}(t^2)$, via a typical application of the matrix inversion lemma; for details, see (\hspace{-0.1pt}\cite{kalogerias2018spatially}, section VI).

\subsubsection{Reduced-Complexity Sample Average Approximation}

 Having determined the necessary posterior statistics involved in \eqref{posterior_1}, our next step would be to evaluate the objective of \eqref{conditional} or, equivalently, the multidimensional integral \eqref{objective}. However, to the best of our knowledge, a closed-form representation of \eqref{objective} is impossible to derive. Therefore, we resort to a near-optimal approach. In particular, we rely on the SAA method, and replace \eqref{conditional} by an easily computable \textit{surrogate}, constructed via \textit{unconditional} Monte Carlo sampling.
 
 To define the proposed surrogate to \eqref{conditional}, fix $\mathbf{p}\in \mathscr{C}_r(t)$ and $t=1,\ldots,N_T$, and consider the change of variables (again, overloading notation)
 \begin{align}\label{changeofvariables}
{v}_{\tau}&=(\sigma_{\tau}^{t+1|t})^{-1}({z}_{\tau}-\mu_{\tau}^{t+1|t}),\,\, \forall \tau\in \mathcal{S}_r^f \cup \mathcal{S}_r^g \,\,\, \text{and} \\
\bm{v}_{\tau_r}^{f,g}&=\big( {\bm{\Sigma}^{t+1|t}_{\tau_r}}(\mathbf{p})\big)^{-1/2} \big(\bm{z}_{\tau_r}^{f,g}-\bm{\mu}_{\tau_r}^{t+1|t}(\mathbf{p})\big),
\end{align}
to the integral of \eqref{objective}. Additionally, also define the collections $\mathcal{V}_r^f\triangleq\{\bm{v}^{f,g}_{\tau_r},\{v_\tau \}_{\tau \in \mathcal{S}_r^f}\}$ and $\mathcal{V}_r^g\triangleq\{\bm{v}^{f,g}_{\tau_r},\{v_\tau \}_{\tau \in \mathcal{S}_r^g}\}$. Then, \eqref{objective} may be equivalently represented as
\begin{align}\label{normalize_SAA}
     \hspace{-2pt}\mathbb{E}[{V}_I(\boldsymbol{S}_r(\mathbf{p},t+1)\hspace{-0.5pt})|\,\mathcal{C}_r(t)]\hspace{-1.5pt}=\hspace{-1.5pt}\int {\bar{V}}^{t+1|t}_I (\mathbf{p}, \boldsymbol{s})   p_{\bar{\boldsymbol{S}}}(\boldsymbol{s}  )  \text{d}\boldsymbol{s}, 
\end{align}
where
\begin{align} \label{theVI_Bar}
   &\hspace{0.2pt} {\bar{V}}^{t+1|t}_I (\mathbf{p},\mathcal{V}_r^f, \varphi^f,\mathcal{V}_r^g, \varphi^g) \\ &\hspace{-2pt}=\hspace{-2.5pt} \frac{P_CP_S\mathsf{F}^{t+1|t}\hspace{-0.5pt}\big(\mathbf{p},\hspace{-0.5pt} \mathcal{V}_r^f , \varphi^f \big) \mathsf{G}^{t+1|t}\hspace{-0.5pt}\big(\mathbf{p},\hspace{-0.5pt} \mathcal{V}_r^g , \varphi^g\big) }{P_S\sigma_D^2\mathsf{F}^{t+1|t}\hspace{-0.5pt}\big(\mathbf{p},\hspace{-0.5pt} \mathcal{V}_r^f\hspace{-0.5pt}, \varphi^f \big)\hspace{-2.5pt}+\hspace{-2.5pt}P_C \sigma^2 \mathsf{G}^{t+1|t}\hspace{-0.5pt}\big(\mathbf{p},\hspace{-0.5pt} \mathcal{V}_r^g\hspace{-0.5pt}, \varphi^g\big)\hspace{-2.5pt}+\hspace{-2pt}\sigma^2\sigma_D^2}\notag
 \end{align}
with the functions $\mathsf{F}^{t+1|t}$ and $\mathsf{G}^{t+1|t}$ being defined as\footnote{If $\boldsymbol{x}$ is a vector, $\boldsymbol{x}|_i$ denotes its $i$-th entry}
\begin{align} \label{theVI_Bar-B}
   &\mathsf{F}^{t+1|t}\hspace{-0.5pt}\big(\mathbf{p},\hspace{-0.5pt} \mathcal{V}_r^f , \varphi^f \big) \hspace{-2pt} \notag \\
   &\quad\triangleq\hspace{-1pt}\mathsf{F}\Big(\hspace{-0.5pt}\mathbf{p}, \big({\bm{\Sigma}^{t+1|t}_{\tau_r}}(\mathbf{p})\big)^{\hspace{-0.5pt}1/2}\bm{v}_{\tau_r}^{f,g} \hspace{-1.5pt}+\hspace{-1.5pt} \bm{\mu}_{\tau_r}^{t+1|t}(\mathbf{p})\big|_{1},\notag \\ 
   \hspace{-2pt}& \quad\quad\quad\{ v_{\tau}\sigma_{\tau}^{t+1|t}\hspace{-1.5pt}+\hspace{-1.5pt}\mu_{\tau}^{t+1|t} \}_{\tau\in \mathcal{S}^f_r},\varphi^f\Big)
\end{align}
and
\begin{align} \label{theVI_Bar-C}
   &\mathsf{G}^{t+1|t}\hspace{-0.5pt}\big(\mathbf{p},\hspace{-0.5pt} \mathcal{V}_r^g , \varphi^g \big) \hspace{-2pt} \notag \\
   &\quad\triangleq\hspace{-1pt}\mathsf{G}\Big(\hspace{-0.5pt}\mathbf{p}, \big({\bm{\Sigma}^{t+1|t}_{\tau_r}}(\mathbf{p})\big)^{\hspace{-0.5pt}1/2}\bm{v}_{\tau_r}^{f,g} \hspace{-1.5pt}+\hspace{-1.5pt} \bm{\mu}_{\tau_r}^{t+1|t}(\mathbf{p})\big|_{2},\notag \\ 
   \hspace{-2pt}& \quad\quad\quad\{ v_{\tau}\sigma_{\tau}^{t+1|t}\hspace{-1.5pt}+\hspace{-1.5pt}\mu_{\tau}^{t+1|t} \}_{\tau\in \mathcal{S}^g_r},\varphi^g\Big),
\end{align}
and where $\bar{\boldsymbol{S}}$ follows the distribution induced by the density
\begin{align} 
    &\hspace{-14pt}p_{\bar{\boldsymbol{S}}}(\mathcal{V}_r^f, \varphi^f,\mathcal{V}_r^g, \varphi^g)\notag\\ &\hspace{-9pt}=
     \mathcal{N}(\bm{v}_{\tau_r}^{f,g};\mathbf{0},\mathbf{I}_2) \hspace{1.5pt}\mathcal{U}({\phi}_{\tau_r}^f; 0,1 )\hspace{1.5pt}\mathcal{U}({\phi}_{\tau_r}^g; 0,1 )\notag\\
    &\times \hspace{-1pt} \prod_{{\tau}\in \mathcal{S}^f_r \cup \mathcal{S}^g_r } \hspace{-1pt}\mathcal{N}({z}_{\tau}; 0,1)  \hspace{1.5pt}\mathcal{U}({\phi}_{\tau}; 0,1 ).\hspace{-13pt} \label{posterior_bar}
\end{align}


The representation \eqref{normalize_SAA} exhibits an important and rather practically appealing property: The density $p_{\bar{\boldsymbol{S}}}$ is \textit{completely independent of both $\mathbf{p}$ and $\mathcal{C}_r(t)$}, and \textit{all} such dependence has been transferred to $\bar{V}_I^{t+1|t}$. Consequently, \textit{sampling} from $p_{\bar{\boldsymbol{S}}}$ is greatly facilitated, and this fact is exactly what makes our proposed SAA-based scheme attractive, whose description now follows.
\begin{algorithm}
 {Joint Beamforming \& Relay Selection Scheme}
\label{alg:SAA_alg}
\begin{algorithmic}[1]
\For{$t=1:N_T$}
\State \underline{\textbf{Beamforming}} ($2$nd stage of time slot $t$)
\State \textbf{Inputs:} Channel aggregates $|\mathbf{1}^{\text{T}}\mathbf{f}_r|^2$ and  $|\mathbf{1}^{\text{T}}\mathbf{g}_r |^2$.
\State Compute optimal $\mathbf{w}_{opt}(t)$ from \eqref{optimalweights}
\State \underline{\textbf{Relay selection}} ($1$st stage of time slot $t+1$)
\For{every cluster $r$}
\State \textbf{Inputs:} a) RSS $\{{z_{\tau}(t)}\}_{\tau \in \mathcal{S}_r^f \cup \mathcal{S}_r^g}$ until $t$.
\State \hspace{30pt} b) RSS ${z^f_{\tau_r}(t)}$ and ${z^g_{\tau_r}(t)}$ until $t$.
\State Generate $\bar{\boldsymbol{S}}$ from  \eqref{posterior_bar}
\For{each $\mathbf{p} \in \mathscr{C}_r(t)$}
\State Compute $\hat{V}_I(\mathbf{p}, t+1)$ from \eqref{SAA}
\EndFor
\State Choose $\mathbf{p}_r(t+1)\in \text{argmax}_{\mathbf{p}\in \mathscr{C}_r(t)}\hat{V}_I(\mathbf{p},t+1)$ 
\EndFor
\EndFor
\end{algorithmic}
\end{algorithm}

For each relay cluster $r$ and at every time slot $t$, the SAA method works by randomly generating a total of $N_S$ \textit{scenarios}, drawn from the distribution  induced by $p_{\bar{\boldsymbol{S}}}$. Clearly, due to the special form of $p_{\bar{\boldsymbol{S}}}$, this is straightforward to implement. Then, each scenario $\bar{\boldsymbol{S}}^{(i)}, i=1,\ldots, N_S$ is used to evaluate $\bar{V}_I^{t+1|t}$, at every possible relay position within the set of feasible locations, $\mathscr{C}_r(t)$. Finally, leveraging \eqref{normalize_SAA}, the SAA of \eqref{objective} is formulated by replacing the expectation in its objective with an \textit{empirical mean} as
\vspace{-2pt}
\begin{align} \label{SAA}
   \begin{split}
    \underset{\mathbf{p}}{\text{maximize}} &\quad \hat{V}_I(\mathbf{p},t+1) \triangleq \frac{1}{N_S}\sum_{i=1}^{N_S} \bar{V}_I^{t+1|t} (\mathbf{p}, \bar{\boldsymbol{S}}^{(i)} )\\
    \text{subject to} &\quad \mathbf{p} \in \mathscr{C}_r(t)
    \end{split},
\end{align}
which may be solved by enumeration. The optimal solution of \eqref{SAA} corresponds to the selected relay at $t+1$.

Note that, following \cite{kalogerias2018spatially}, it may be argued that exactly the same set of scenarios may be used by all relays, in all clusters, and even at all time slots. This, of course, keeps the sampling requirements at a bare minimum, \textit{networkwide}.

\begin{figure}
    \centering
    \includegraphics[scale=1.08]{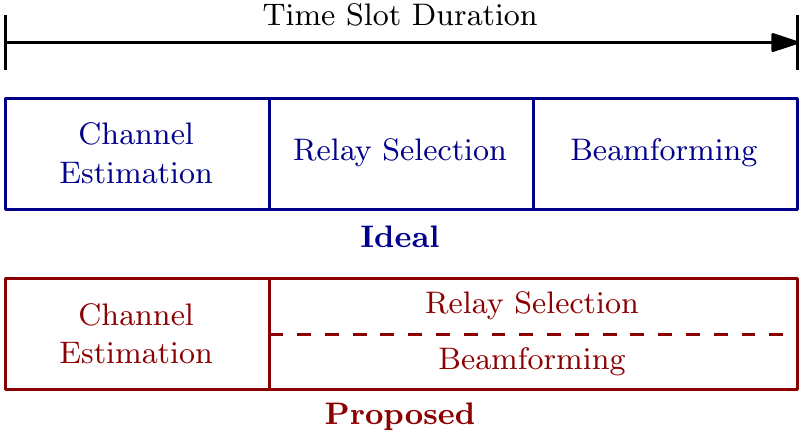}
    \caption{The required phases executed per cluster in the duration of the time slot for the ideal and proposed schemes. The dashed line indicates that the relay selection and beamforming tasks are performed in parallel, resulting in a better utilization of the time slot for the communication task}
    \label{fig:timeslot}
\end{figure}


\subsection{2-stage joint beamforming/relay selection}

Our $2$-stage joint beamforming and relay selection scheme is described in Algorithm \ref{alg:SAA_alg}. At time $t$, beamforming towards the destination is performed, which corresponds to the $2$nd \textit{stage problem} of time slot $t$. In this stage, the RSS and phases of the channel aggregates at every cluster representative need to be centrally collected, in order to compute the optimal beamforming weights. Within the same time slot $t$, and in parallel to beamforming, the $1$st \textit{stage problem} of time slot $t+1$ is solved, i.e., every cluster individually selects the relay to be used for beamforming in the subsequent time slot. In this stage, in addition to the CSI of the cluster representative at $\mathbf{p}_r$, the relay selection process also requires the CSI of the unique segments that comprise the propagation paths to that cluster. This information can be easily acquired via low cost devices, e.g., channel sounders, placed on every street segment, and then sent through the backhaul network to the respective cluster.  

In practice, it is sufficient to condition on a window of past time slots, as opposed to the entire observed RSS history. Such an approximation is expected to work well even for a relatively small window size, due to the exponentially decaying structure of the temporal correlation component of the channel model. Moreover, depending on the mmWave channel coherence time, it might be sufficient to follow a two-timescale design, where the beamforming weights would be computed in every time slot but relay selection would be executed over a longer time interval.

\begin{figure*}[t] 
\vspace{-6pt}
\centering
\subfloat[]{\label{fig:Allpolicies2}\includegraphics[width=0.32\textwidth]{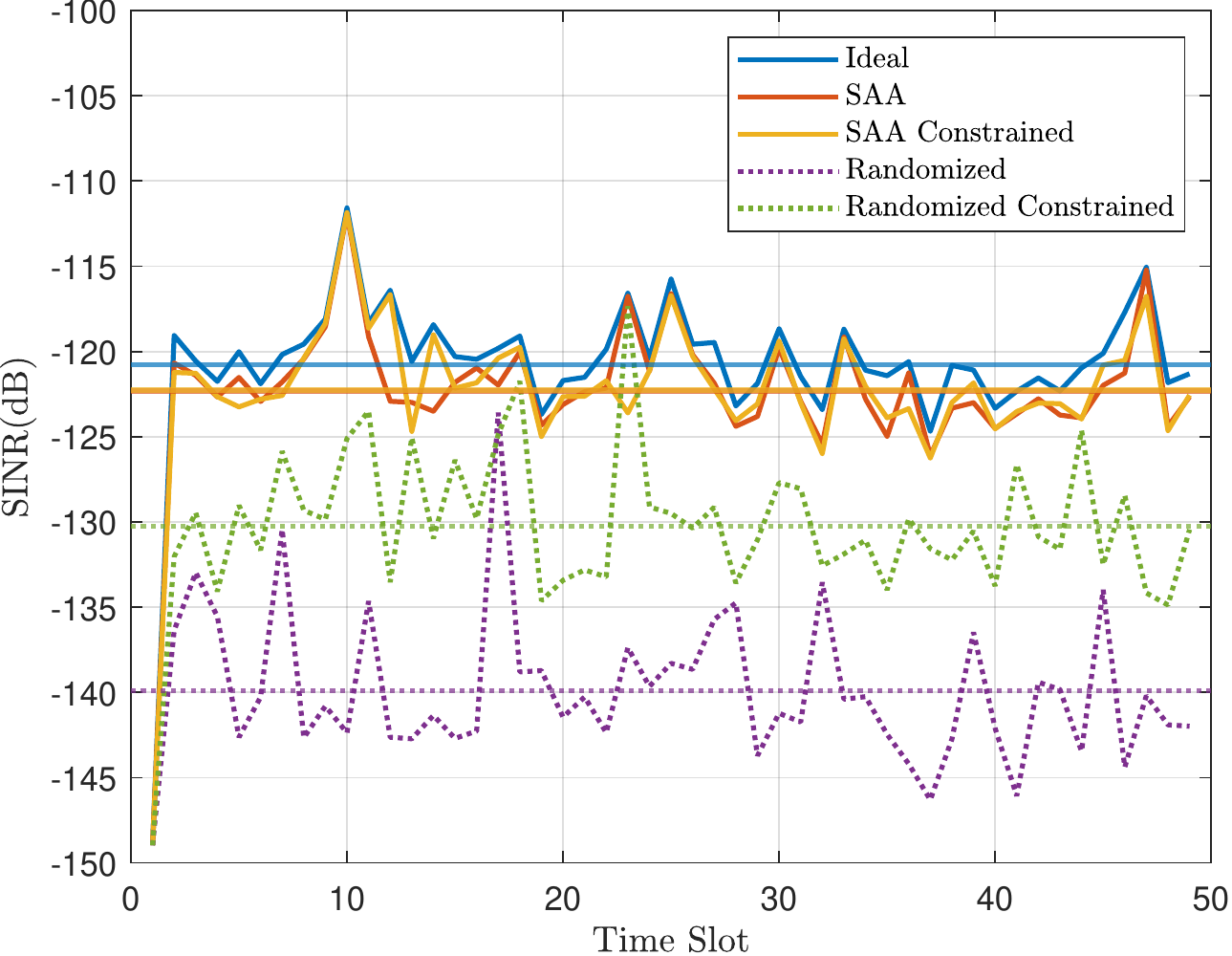} }%
\hfil
\subfloat[]{\label{fig:Allpolicies4}\includegraphics[width=0.32\textwidth]{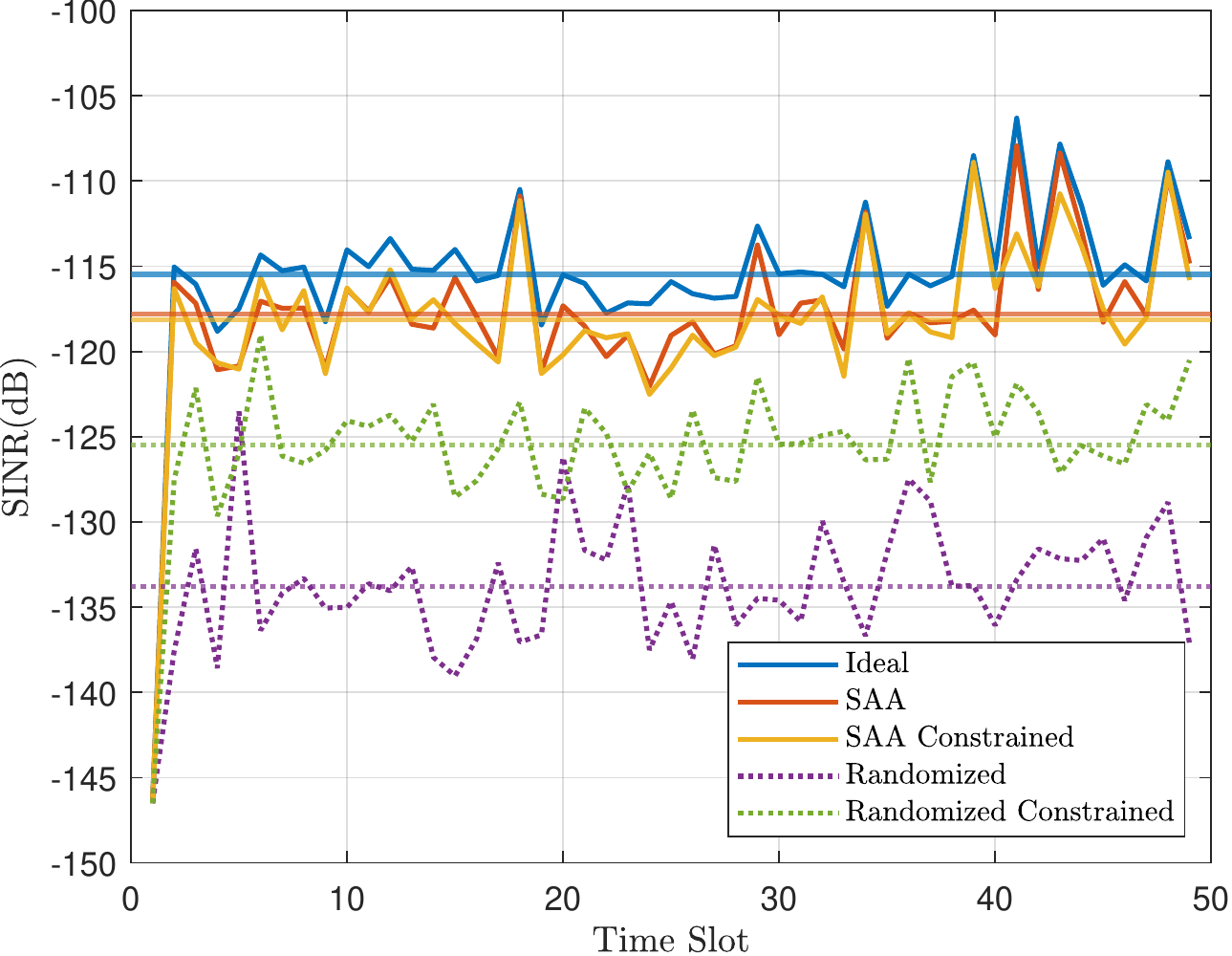} }%
\hfil
\subfloat[]{\label{fig:Allpolicies6}\includegraphics[width=0.32\textwidth]{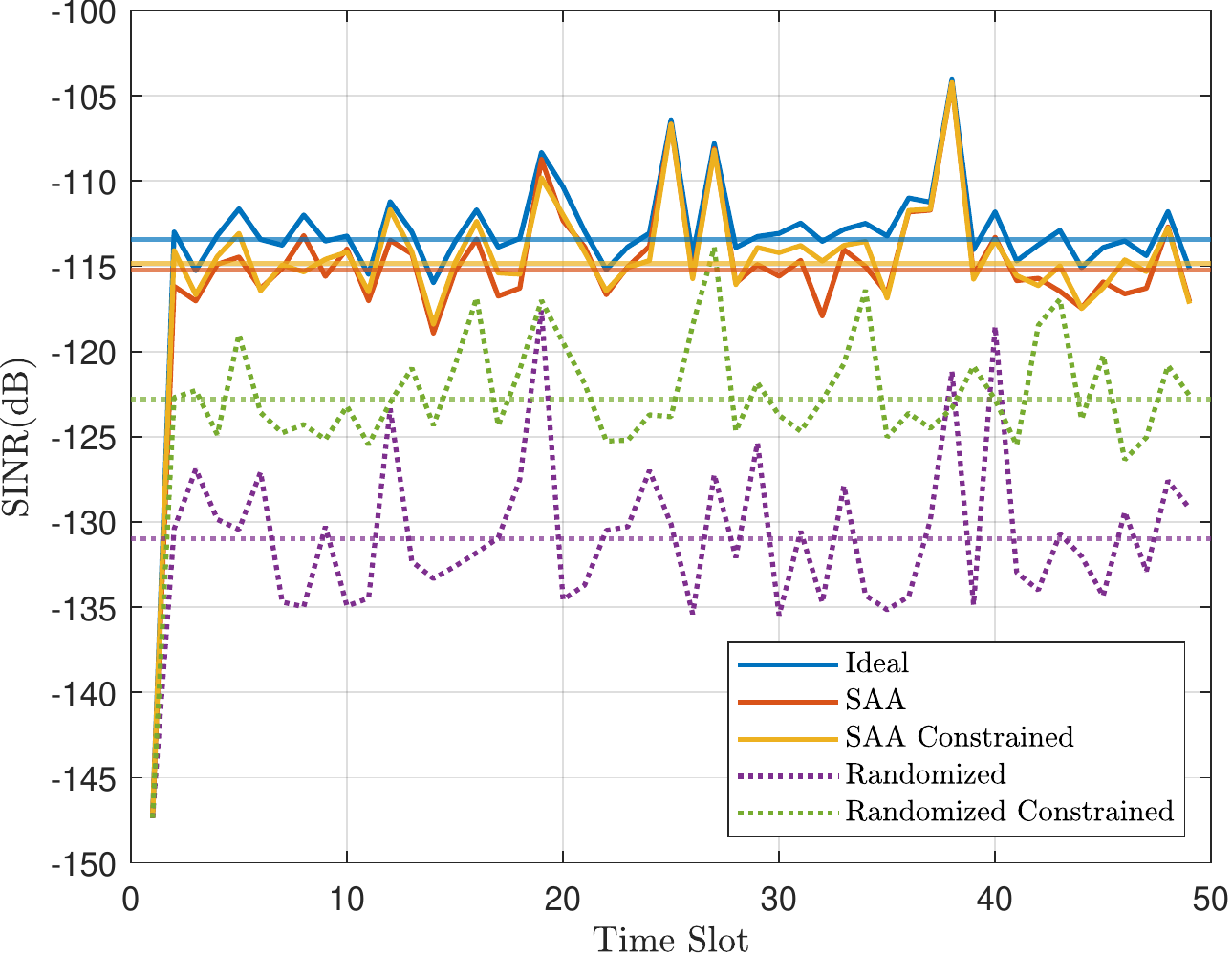} } %
 \caption{The average SINR observed in every time slot over $10000$ independent trial runs, for all policies, when keeping only (a) two (b) four (c) six clusters of Fig. \ref{topology}. The horizontal lines indicate the average SINR across all time slots for that respective policy.}
\end{figure*}

\section{Operational phases of the time slot} \label{timeslotphase}
In this section, we discuss how the operations of relay selection and beamforming are scheduled within each time slot, comparatively for the proposed and ideal selection schemes.  

In every time slot of the ideal scheme (Fig. \ref{fig:timeslot} top), relay selection is always implemented \textit{before} optimal beamforming; this is simply due to the fact that acquisition of the \textit{current} RSS has to inevitably be performed in the same time slot as beamforming. On the other hand, in the proposed scheme (Fig. \ref{fig:timeslot} bottom), relay selection at the current time slot is implemented \textit{predictively during the previous time slot}, by efficiently exploiting past RSS observations. As a result, the overhead caused by the relay selection process can be effectively bypassed, and optimal beamforming at the current time slot may be implemented \textit{completely in parallel with} the predictive relay selection affecting the next time slot. 


Next, let us look at the CSI estimation requirement of each selection scheme into more detail. In the ideal scheme, the incoming and outgoing channels of every relay for all clusters are initially estimated. This requires estimating $N_{ideal}= 2\delta N_c$ distinct channels. Channel estimation is initiated by a pilot symbol broadcaster from the source, with every relay of all clusters measuring their RSS. A similar procedure is done for estimating the respective channels towards the destination. On the contrary, our proposed scheme requires estimating only the CSI of the cluster representatives, as well as the CSI of segment $\tau\in \mathcal{S}_{r}^f \cup \mathcal{S}_{r}^g$, for all $r=1,\ldots,N_c$. Therefore, $N_{proposed}=2N_c +|\cup_{r=1}^{N_c}\mathcal{S}^f_r \cup \mathcal{S}^g_r|$ channels have to be estimated, where $|\cdot|$ denotes the cardinality of a set. 

Compared to the ideal, the proposed relay selection scheme is particularly advantageous in dense network topologies, where, to account for high channel variability, each cluster needs to include a large number of relays, and the number of relays per cluster is relatively larger than the number of segments taking part in the communication. Actually, a dense network is required even if the shadowing variance is low, since this implies weaker channel correlation, due to the now dominant multi-path fading. It is then clear that our proposed scheme requires significantly fewer channels to be estimated, which in turn leads to reduced channel estimation overhead. 

Note that, while our proposed scheme does incur a computational burden associated with the relay selection process, due to the need for execution of Algorithm \ref{alg:SAA_alg}, the parallelization of relay selection and optimal beamforming in each time slot not only compensates for that burden, but also naturally leads to more consistent ergodic performance, as long as the accuracy of predictive relay selection is adequate.
 


\section{Simulations} \label{sect:simulations}
We examine the performance of the proposed relay selection scheme using synthetically generated CSI data. For our simulations, we assume the topology and cluster placement of Fig. \ref{topology}, unless otherwise stated. All segments are of length $d_{full}=100m$. For  segments containing a relay cluster, we consider $\delta=50$ evenly spaced positions upon which the individual relays are located. We assume a carrier frequency of $28$ GHz, with path-loss exponent similar to those reported in  \cite{maccartney2014omnidirectional,park2016millimeter}, i.e. $\alpha_L=\alpha_N=2.1$. The assumed channel parameters are $\eta^2=40$, $\gamma=15$, $\beta=10m$, $\Delta=10$dB, $\sigma^2_{\xi}=20$, $\sigma^2=1$, and $\sigma^2_D=1$. The transmission power (which also includes the directional antenna gain) is $P_S=80$dBm and the total relay power budget is $P_C=100$dBm. The relay clusters are connected to a stable power source.

To assess the long term system performance of our SAA-based system, we compare it against two benchmark relay selection policies. The first is the ideal policy, where, as described previously, the best relay from every cluster is selected and used in beamforming, during the current time slot. This policy provides an upper bound on the performance of any admissible policy. The second benchmark is a non-strategic, randomized policy, where a relay is chosen uniformly at random from its cluster to be used in the subsequent time slot, irrespective of the observed CSI.
For the SAA and randomized policies, we also examine the constrained neighborhood case, where the only candidate relays are those in the $4$ positions closest to either one of the two cluster segment edges.

All selection policies were executed over $N_T=50$ time slots, with $N_S=500$ SAA scenarios. This constitutes one trial of the whole communication task. The QoS achieved in each time slot, averaged over $10^4$ trials, is shown for all policies in Fig. \ref{fig:Allpolicies4}. Note that assessing the performance of the proposed relay selection policies via averaging is correct and technically justified; see \cite{kalogerias2018spatially} for details.
The superiority of the SAA policy is evident, as it achieves almost $7.7$dB larger SINR than the randomized policy, while also performing only $2.3$dB worse than the ideal policy. This significant result implies that near optimal selection is achieved by exploiting one-step ahead SINR prediction, rather than waiting for the actual future channel values, available at the next time slot.

\begin{figure*}[t] 
\vspace{-6pt}
\centering
\subfloat[]{\label{fig:Ideal_R1}\includegraphics[width=0.24\textwidth]{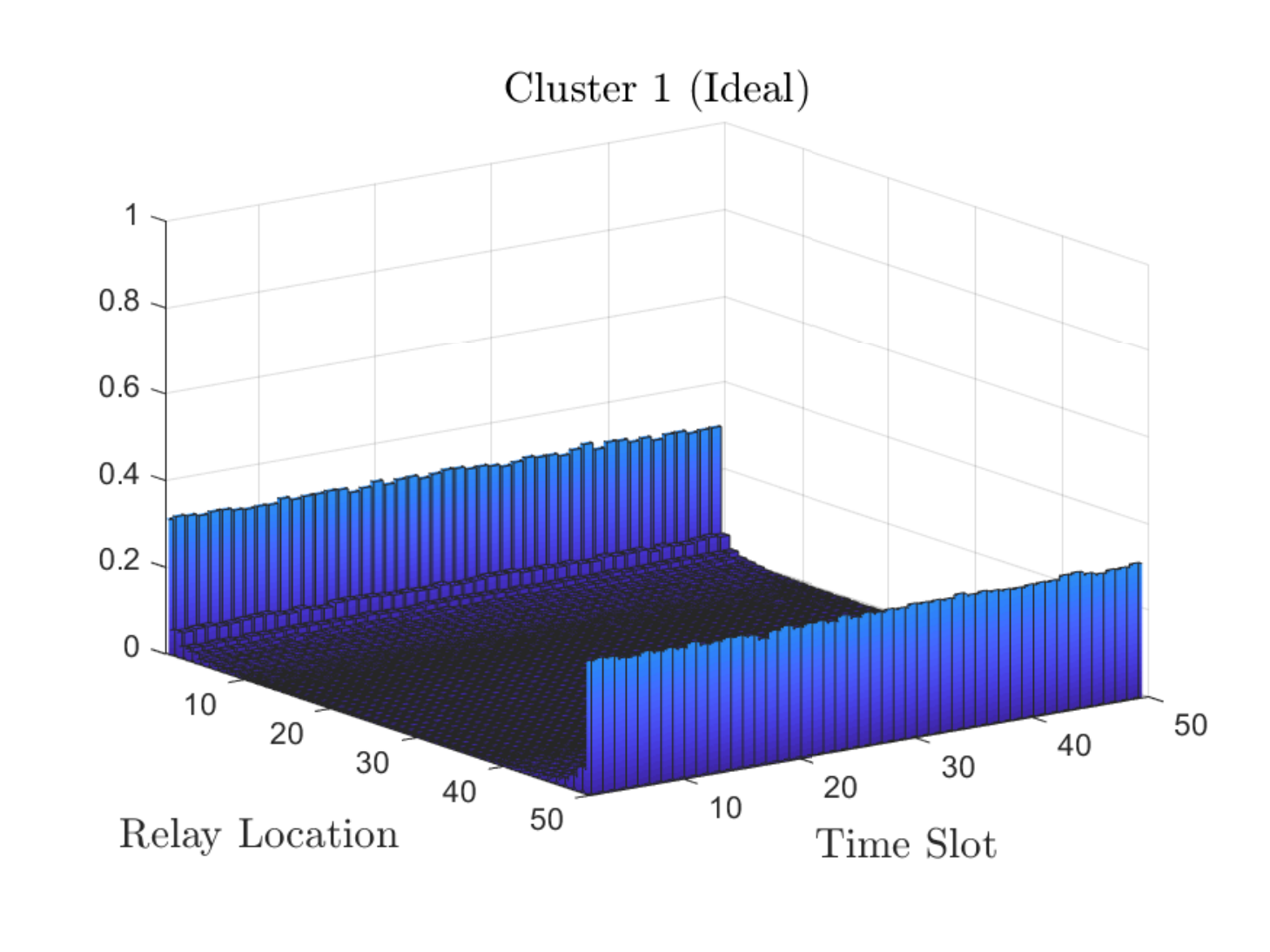} }%
\subfloat[]{\label{fig:Ideal_R2}\includegraphics[width=0.24\textwidth]{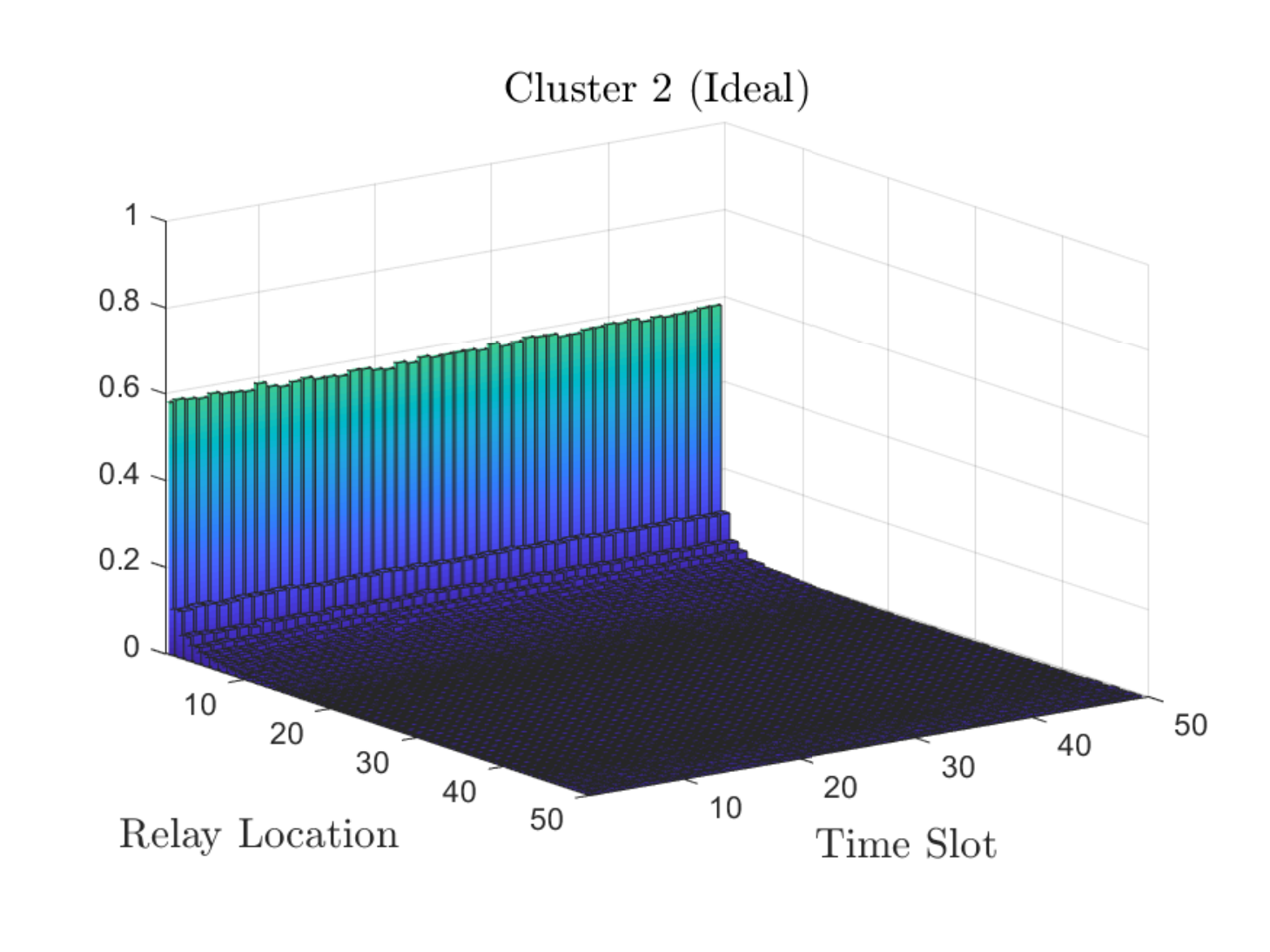} }%
\subfloat[]{\label{fig:Ideal_R3}\includegraphics[width=0.24\textwidth]{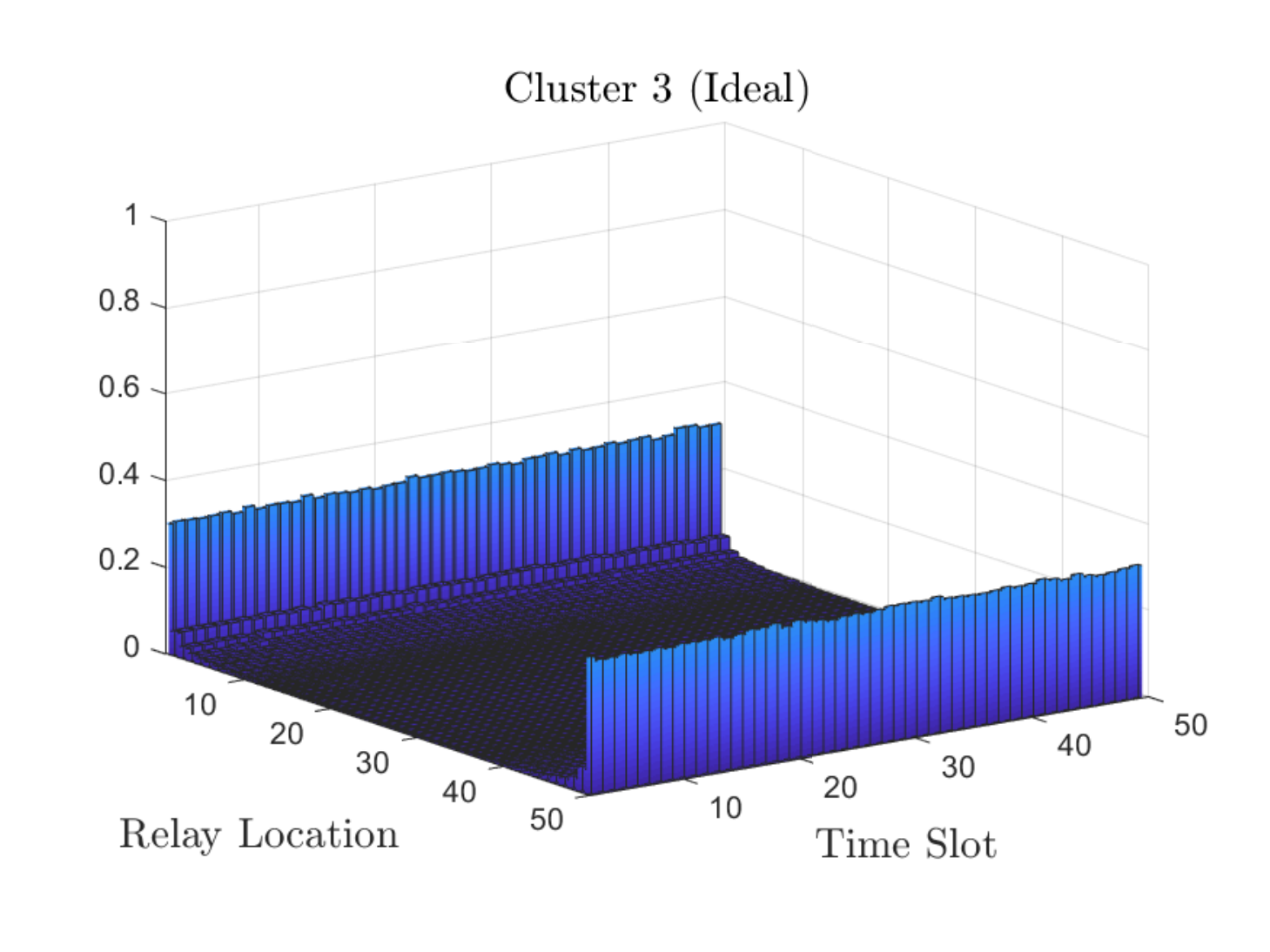} }%
\subfloat[]{\label{fig:Ideal_R4}\includegraphics[width=0.24\textwidth]{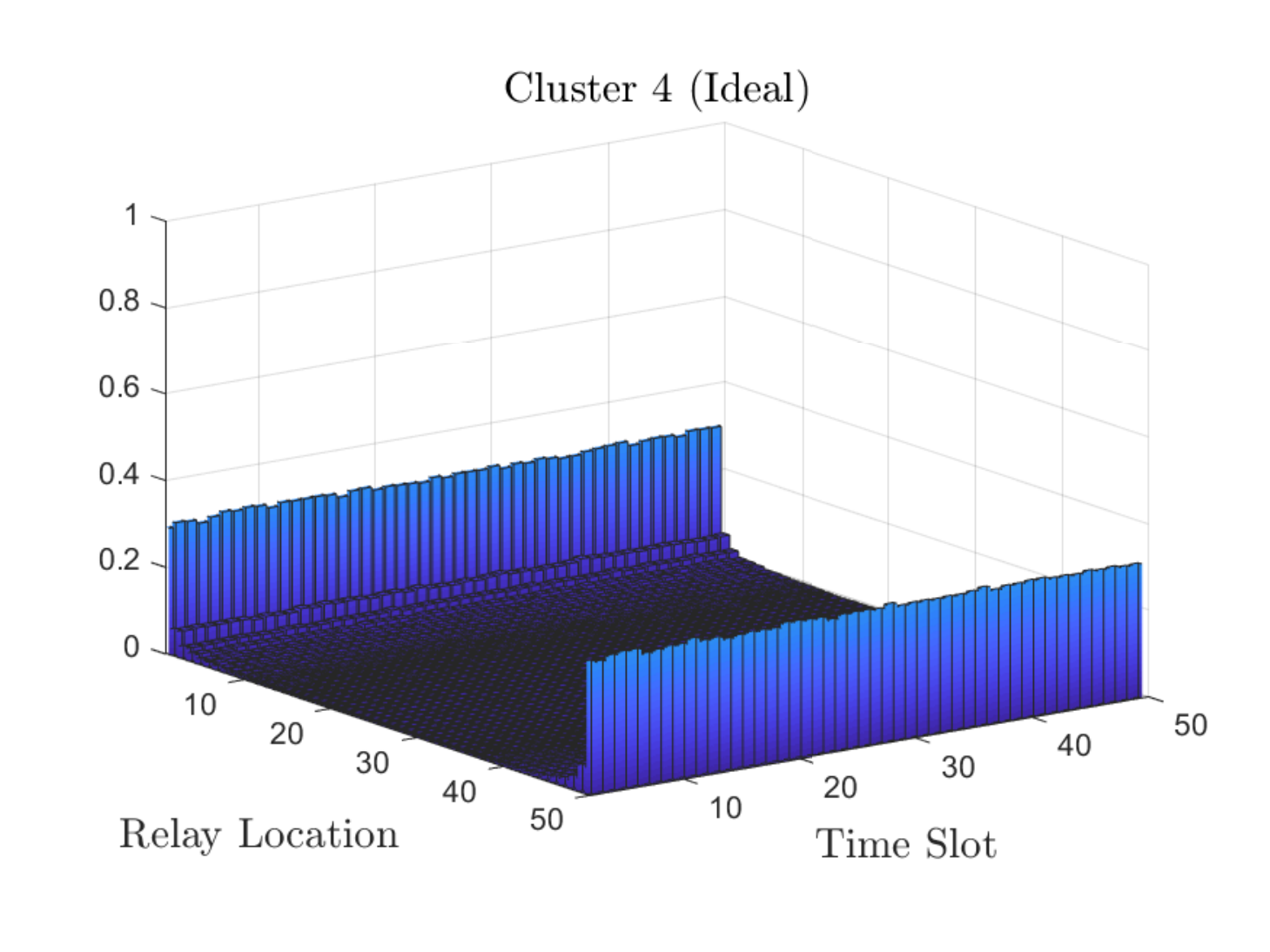} }%
 \caption{Ideal policy: fraction of times over $10000$ independent trials that a relay was chosen in each time slot, for relay clusters (a) 1 (b) 2 (c) 3 (d) 4 of Fig. \ref{topology}. For each trial the same relay is initially selected in all clusters, so time slot $t=1$ is omitted from the histograms.}
\label{fig_histograms_ideal}
\end{figure*}

For the ideal and SAA approach, we also show the histograms of the fraction of times a specific relay was chosen in every time slot. Observing the ideal policy of Fig. \ref{fig_histograms_ideal}, it seems that there is a tendency to select relays located near either of the two end points of the cluster, which for relay clusters $2$ and $3$ is done about $60\%$ of the time. This is more evident in the SAA approach where at every time slot about $99\%$ of the total trials choose either of the end point relays.

Under the same channel and simulation parameters, we varied the number of available clusters. We first considered the case where the topology consisted only of clusters $1$ and $2$ of Fig. \ref{topology}, and another were in addition to the four clusters of Fig. \ref{topology}, two more were placed between intersections $6$ and $5$, and $4$ and $7$. The average SINR for all policies of the two and six cluster configurations are plotted in Fig. \ref{fig:Allpolicies2} and \ref{fig:Allpolicies6}, respectively. Notice that, as compared to Fig. \ref{fig:Allpolicies4}, the average SINR achieved for the two cluster configuration performs about $5$dB worse for all policies, while for the six cluster configuration the performace is about $3$dB better. This implies that the more relay clusters present in the city, the better the overall SINR performance achieved at the destination. Of course, when more relays are involved the synchronization overhead increases.

\begin{figure}[ht] 
\centering
\subfloat[]{\label{topology4other}\includegraphics[width=\linewidth]{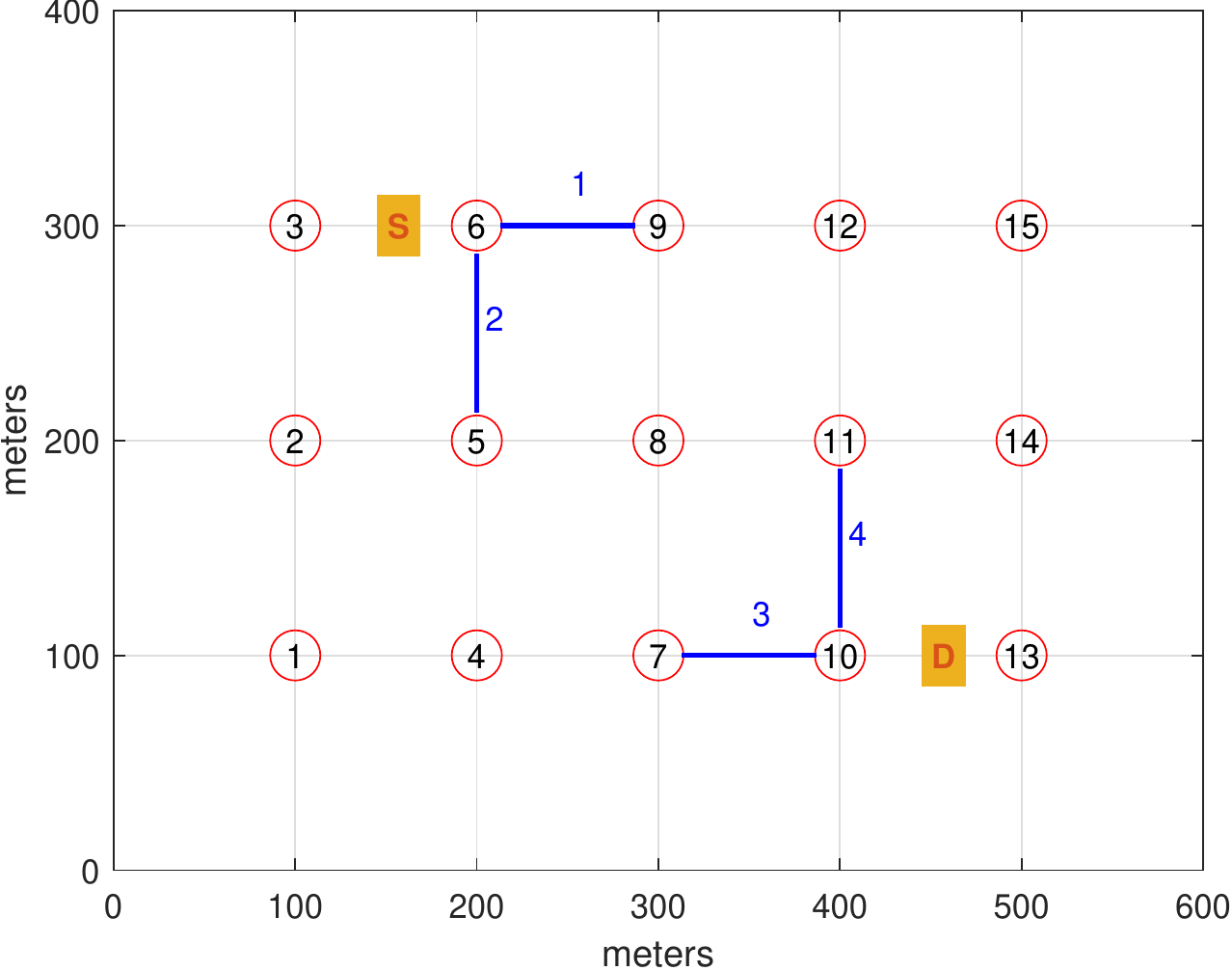}}\\
\hfil
\subfloat[]{\label{allpolicies4other}\includegraphics[width=\linewidth]{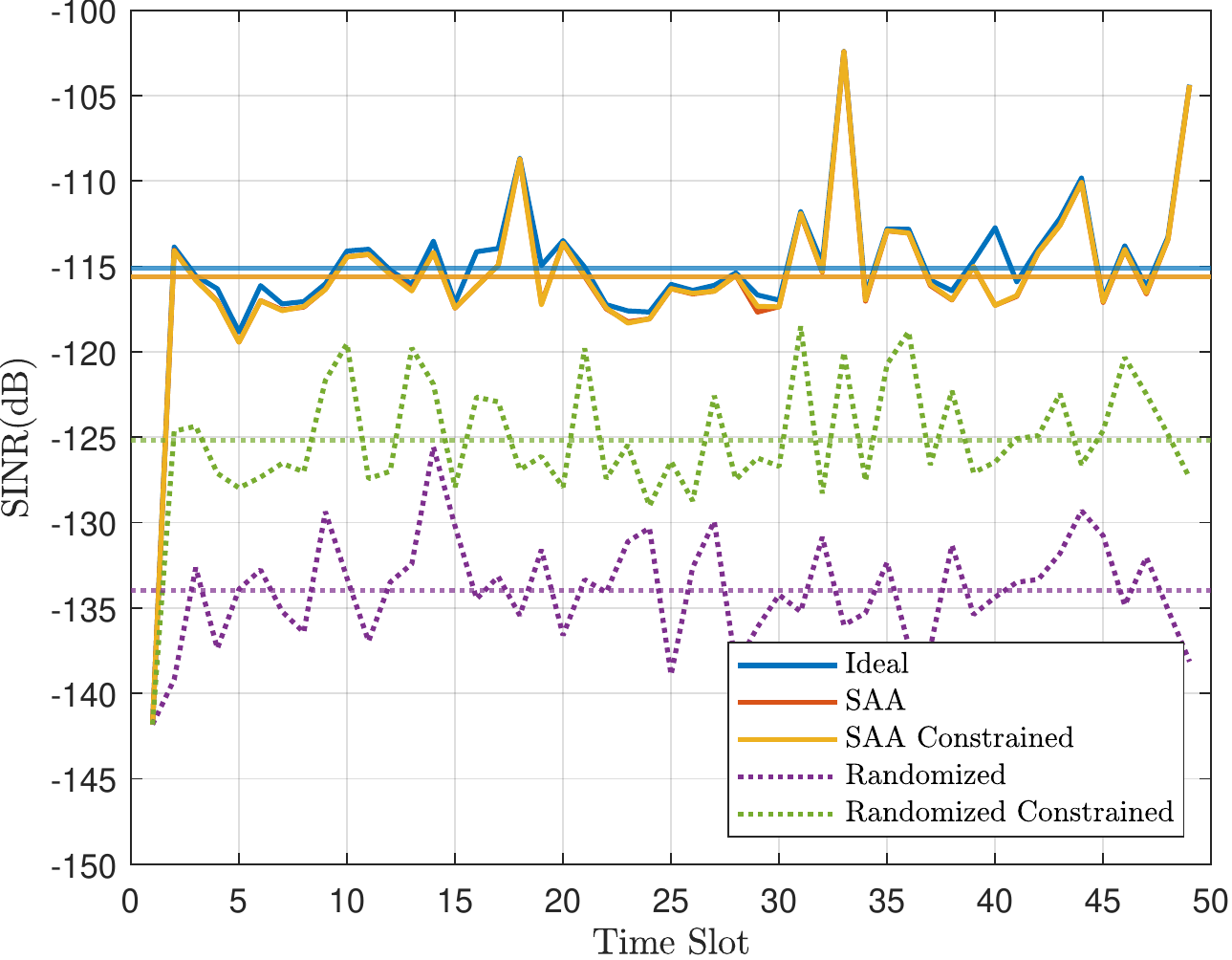}}%
 \caption{(a) A different four cluster topology. (b) The average SINR observed in every time slot for all policies, over $10000$ independent trial runs for the above cluster topology. The horizontal lines indicate the average SINR across all time slots for that respective policy.}
\label{fig:othertopology}
\end{figure}

In Fig. \ref{topology4other} we also examined a topology where two clusters are placed near the source and two near the destination. This case can be thought of as directly assisting the source and destination, fairly. In reality, the probability of this happening in a city is expected to be low as there will be multiple users the clusters would need to service. Still, one can observe that the SINR performance is practically equivalent to that of Fig. \ref{fig:Allpolicies6}. In other words, the cluster placement problem is equally important to the performance of the overall system, and therefore deserves further investigation.




\begin{figure*}[t] 
\vspace{-6pt}
\centering
\subfloat[]{\label{fig:SAA_R1}\includegraphics[width=0.24\textwidth]{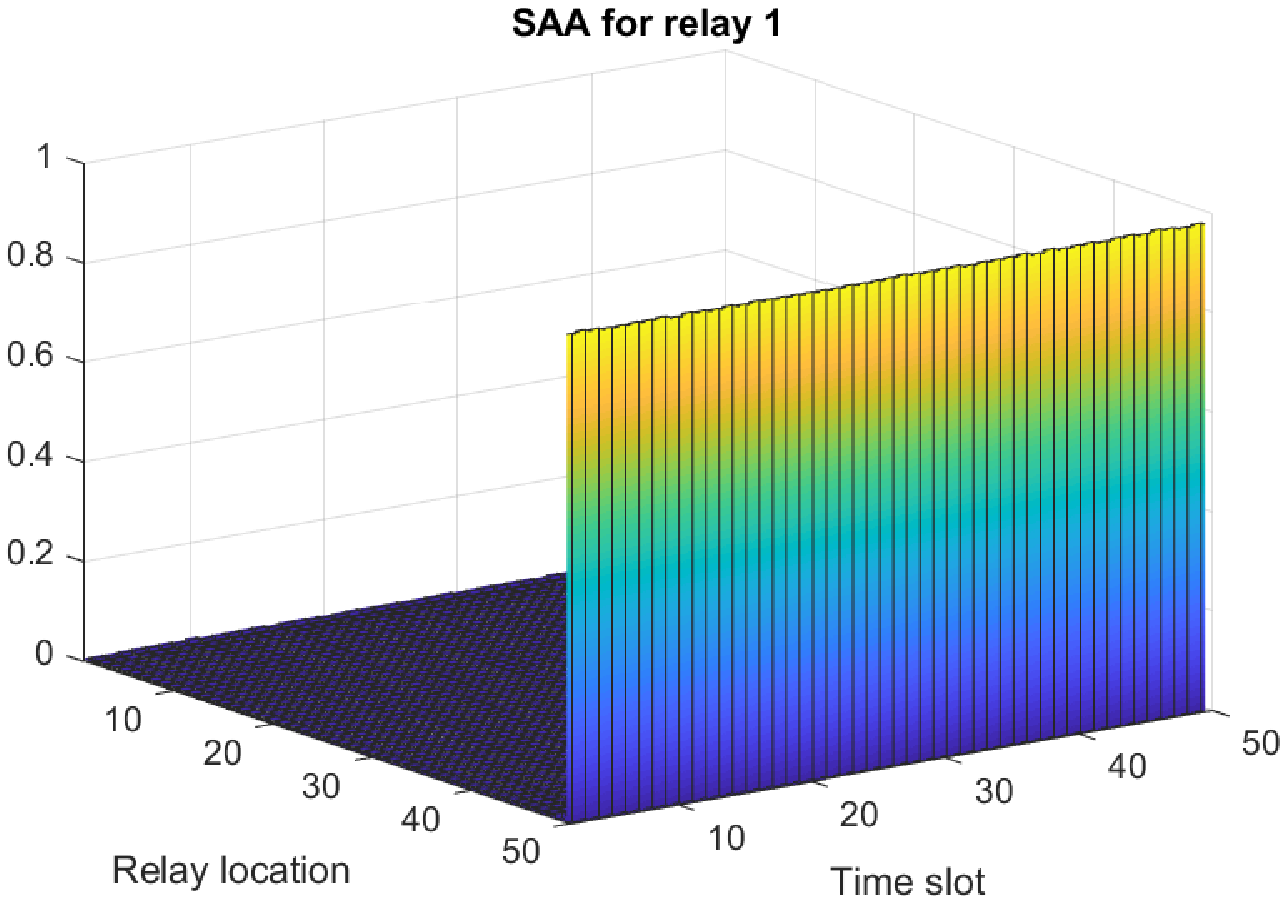}}%
\subfloat[]{\label{fig:SAA_R2}\includegraphics[width=0.24\textwidth]{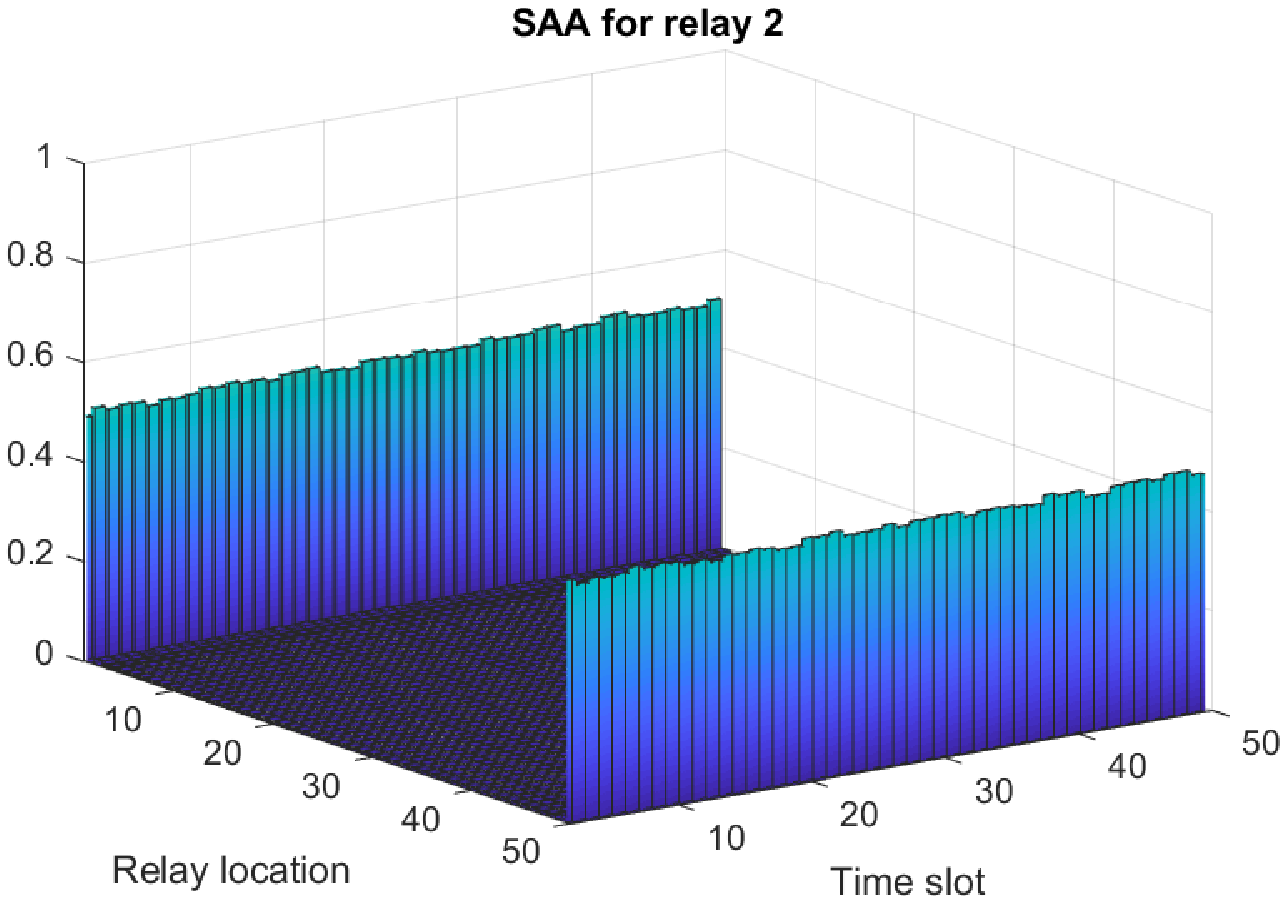} }%
\subfloat[]{\label{fig:SAA_R3}\includegraphics[width=0.24\textwidth]{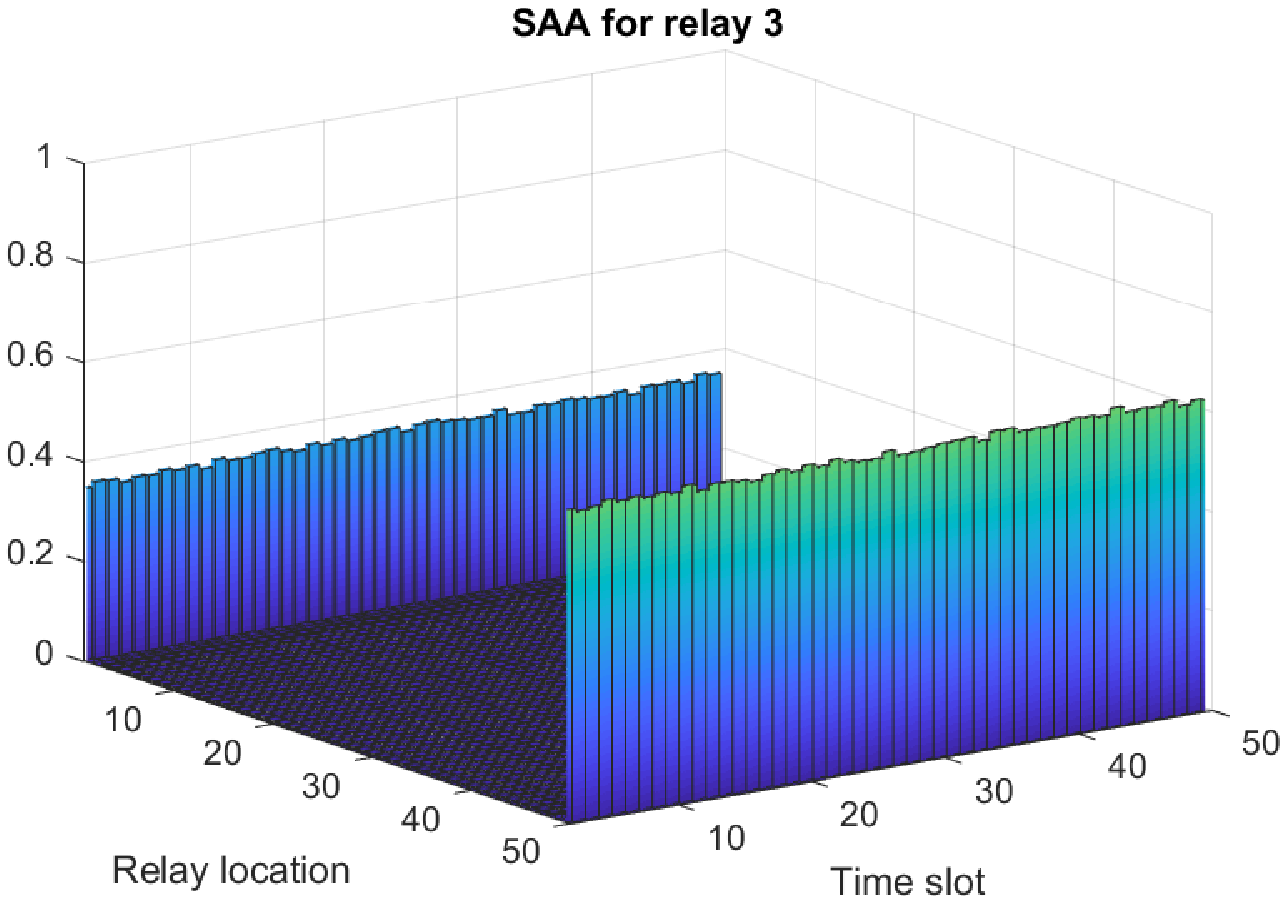} }%
\subfloat[]{\label{fig:SAA_R4}\includegraphics[width=0.24\textwidth]{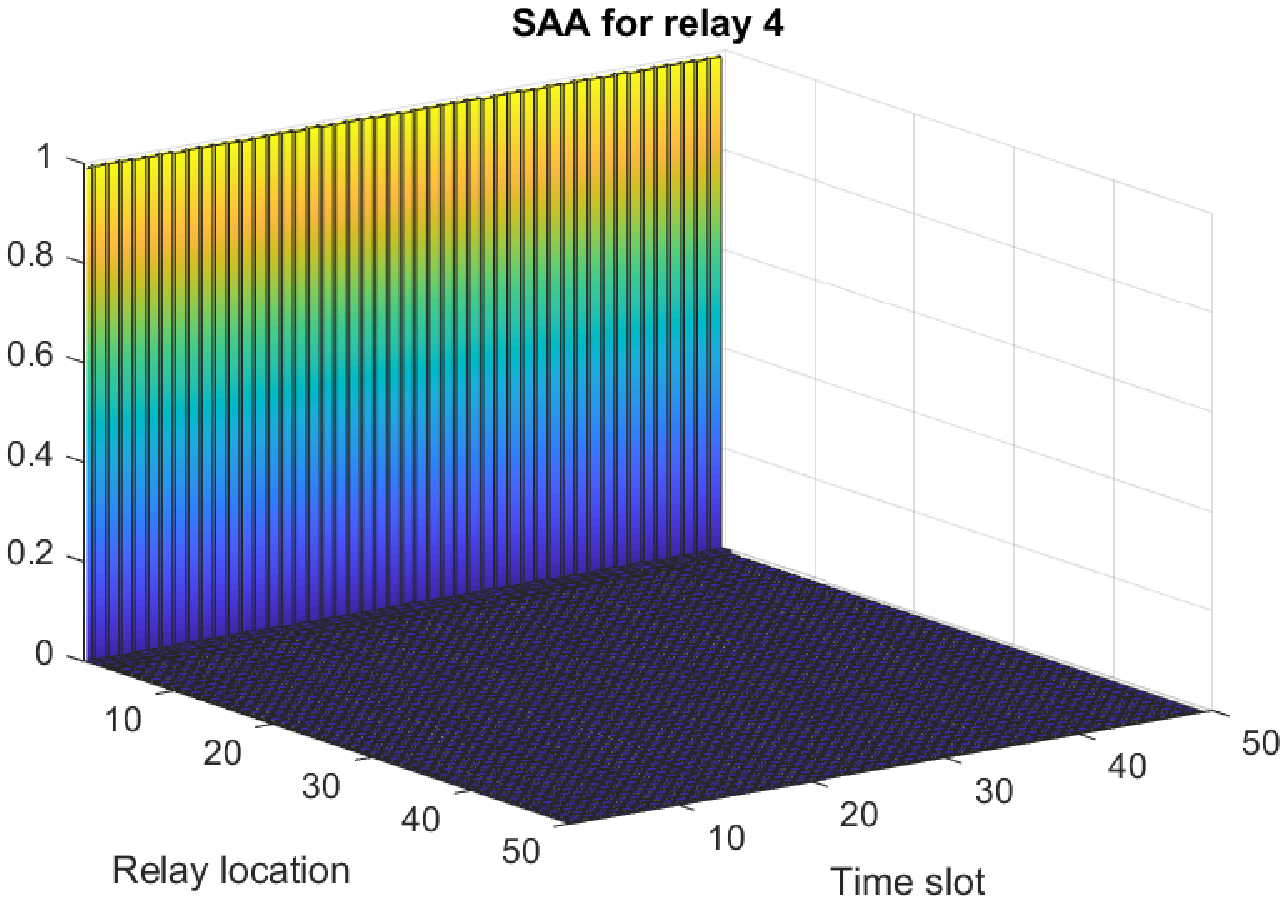} }%
 \caption{SAA policy: fraction of times over $10000$ independent trials that a relay was chosen in each time slot, for relay clusters (a) 1 (b) 2 (c) 3 (d) 4 of Fig. \ref{topology}. For each trial the same relay is initially selected in all clusters, so time slot $t=1$ is omitted from the histograms.}
\label{fig_histograms_SAA}
\end{figure*}

\section{Conclusion} \label{sect:conclusion}
This paper considered a set of spatially distributed clusters of static relays,  cooperatively enhancing mmWave communications in an urban environment. In particular, on a per time slot basis, one relay from each cluster is selected to participate in optimal transmit beamforming towards the destination. In order to decide on the best relay configuration to be used for beamforming, we proposed an adaptive, sampling-based, distributed across clusters and of reduced complexity relay selection scheme, which efficiently exploits spatiotemporal correlations of the mmWave medium, and can be executed completely in parallel to optimal beamforming-based communication. The effectiveness of our approach is established via numerical simulations, which corroborate its superiority against a RSS-agnostic, purely randomized relay selection policy, as well as its efficiency relative to a reference, resource demanding ideal selection scheme. 
As a byproduct of our results, we also demonstrated the sensitivity of the proposed system with respect to spatial cluster placement, confirming that efficient spatial cluster assignment constitutes an interesting, natural direction for future research. 


\appendices
\section*{Appendix}
\label{appendix:b}
If we assume the set $\mathcal{S}_r^f=\cup_{i=1}^{L_r} \mathcal{S}_{ri}^f$ that contains all unique segments of all channels $f_i$ leading to relay $r$, as well as the sets $\mathcal{Z}_r^f=\{z_\tau \}$, $\tau \in \mathcal{S}_r^f$ and set $\varphi^f=\{\phi_\tau\}$, we have that:
\begin{align*}
    &\mathsf{F}\big(\mathcal{Z}_r^f, \varphi^f \big)=  |\mathbf{1}^{\text{T}}\mathbf{f}_r|^2\\ 
    &= |f_{r1}|^2+ \ldots +|f_{rL_r}|^2+ f_{r1}f_{r2}^*+f_{r1}f_{r3}^*+.... \\
    &=  |f_{r1}|^2+ \ldots +|f_{rL_r}|^2+ |f_{r1}|e^{j\Phi_{r1}}|f_{r2}|e^{-j\Phi_{r2}}+... \\
    &= \sum_{i=1}^{L_r}|f_{ri}|^2+|f_{r1}||f_{r2}|(e^{j(\Phi_{r1}-\Phi_{r2})}+e^{j(\Phi_{r2}-\Phi_{r1})})+\ldots \\
    &=\sum_{i=1}^{L_r} \sum_{k=1}^{L_r} e^{\frac{\chi}{2}F_{ri}}e^{\frac{\chi}{2}F_{rk}}\cos(\Phi^f_{ri}-\Phi^f_{rk})\\
     &=\sum_{i=1}^{L_r} \sum_{k=1}^{L_r} \exp\hspace{-2pt}\bigg[{\frac{\chi}{2} \bigg(a^f_{ri}(\mathbf{p}_i)\hspace{-1pt}+\hspace{-1pt}z^f_{\tau_r}(\mathbf{p}_i,t)\hspace{-1pt}+\hspace{-1pt}\sum\nolimits_{{\tau}\in \mathcal{S}_{ri}^f}\hspace{-2pt} z_{\tau}(t)\hspace{-1pt}\bigg)}\hspace{-1pt}\bigg]\\
      &\hspace{20pt}\times \exp\hspace{-2pt}\bigg[{\frac{\chi}{2} \bigg(a^f_{rk}(\mathbf{p}_k)\hspace{-1pt}+\hspace{-1pt}z^f_{\tau_r}(\mathbf{p}_k,t)\hspace{-1pt}+\hspace{-1pt}\sum\nolimits_{{\tau}\in \mathcal{S}_{rk}^f}\hspace{-5pt} z_{\tau}(t)\hspace{-1pt}\bigg)}\hspace{-1pt}\bigg] \hspace{-6pt} \\
     &\hspace{25pt}\times \cos\hspace{-2pt}\bigg(\phi^f_{\tau_r}(\mathbf{p}_i,t)\hspace{-1pt}-\hspace{-1pt}\phi^f_{\tau_r}(\mathbf{p}_k,t)\hspace{-1pt}\\
     &\hspace{60pt}+\hspace{-1pt}\sum\nolimits_{{\tau}\in \mathcal{S}_{ri}^f}\hspace{-3pt}\phi_{\tau}(t)\hspace{-1pt}-\hspace{-1pt}\sum\nolimits_{{\tau}\in \mathcal{S}_{rk}^f}\phi_{\tau}(t)\hspace{-1pt}\bigg)
\end{align*}
where $\chi=ln(10)/10$. Note that in the above, the left hand side refers to the unique segments traversed in all paths, while the equation on the right hand side to the ordered segments in a path. A similar expansion holds for $\mathsf{G}(\mathcal{Z}^g , \varphi^g)$.



\bibliographystyle{IEEEbib.bst}
\bibliography{refs.bib}





\end{document}